\documentclass[12pt]{article}
\usepackage{amsmath,amsfonts,amssymb,amsthm}
\usepackage{latexsym}

\font\notefont=cmsl8
\newcommand{\field}[1]{\mathbb{#1}}

\newcommand{\R}{\field{R}}
\newcommand{\C}{\field{C}}
\newcommand{\F}{\mathcal{F}} 
\renewcommand{\H}{\mathcal{H}} 

\newcommand{\ad}{\mbox{ad}}
\newcommand{\eps}{\varepsilon}
\newcommand{\ph}{\varphi}
\newcommand{\const}{\mbox{const}}
\newcommand{\supp}{\mbox{supp}}

\newcommand{\expect}[1]{\mbox{$\langle #1 \rangle $}}

\newcommand{\sprod}[2]{\langle #1,#2 \rangle}

\newtheorem{theorem}{Theorem}
\newtheorem{lemma}[theorem]{Lemma}
\newtheorem{corollary}[theorem]{Corollary}

\newtheorem{proposition}[theorem]{Proposition}

\newcommand{\Ran}{\operatorname{Ran}}
\newcommand{\Ima}{\operatorname{Im}}

\begin{document}
\title{\bf{Asymptotic Electromagnetic Fields in Models of Quantum-Mechanical
  Matter Interacting with the Quantized Radiation Field}}
\author{\vspace{5pt} J. Fr\"ohlich$^1$\footnote{juerg@itp.phys.ethz.ch} ,
M. Griesemer$^2$\footnote{marcel@math.uab.edu}
and B. Schlein$^{1}$\footnote{schlein@itp.phys.ethz.ch} \\
\vspace{-4pt}\small{$1.$ Theoretical Physics, ETH--H\"onggerberg,} \\
\small{CH--8093 Z\"urich, Switzerland}\\
\vspace{-4pt}\small{$2.$ Department of Mathematics, University of Alabama at
Birmingham,} \\
\small{Birmingham, AL 35294}\\ }
\date{\small 18 September, 2000}

\maketitle

\begin{abstract}
  In models of (non-relativistic and pseudo-relativistic) electrons
  interacting with static nuclei and with the (ultraviolet-cutoff) quantized
  radiation field, the existence of asymptotic electromagnetic fields is
  established. Our results yield some mathematically rigorous
  understanding of Rayleigh scattering and
  of the phenomenon of relaxation of isolated atoms to their ground states. Our proofs are based
  on propagation estimates for electrons inspired by similar estimates known
  from $N$-body scattering theory.
\end{abstract}

\section{Introduction}
\label{sec:intro}

In this paper we study the scattering of light at non-relativistic and
pseudo-relativistic, quantum mechanical electrons moving under the
influence of an external potential and minimally coupled to the soft modes of
the quantized electromagnetic field. The external potential may be the Coulomb
potential generated by a configuration of static nuclei. Our goal is to
establish the {\em existence of asymptotic electromagnetic fields} on states of
the system with the property that the velocities of all electrons present are smaller
than the velocity of light, (in a sense to be made mathematically precise).
This property is automatically satisfied if one chooses {\em relativistic
kinematics} in the description of electrons, because the propagation velocity
of a {\em massive} relativistic particle is smaller than the velocity of
light. In contrast, if the kinematics of electrons is non-relativistic these
particles can propagate arbitrarily fast, and the condition that the propagation
velocities of electrons in a state of the system are smaller than the velocity of
light is a very stringent one. It is satisfied if all electrons remain bound
to nuclei. But, in realistic models, such ``bound states'' are {\em not dense} in
the Hilbert space of the system.

The key physical idea underlying our analysis is very clear and simple: {\em
Huygens' principle} for the electromagnetic field implies that, on states
with the property that the propagation speeds of all charged particles
are smaller than the velocity of light, the strength of interactions between
the charged particles and the electromagnetic field tends to $0$, as time
$t$ tends to $\pm\infty$, at an integrable rate. As a consequence, one can
use a variant of {\em Cook's method} \cite{C} to prove existence of asymptotic electromagnetic
fields; (a (strong) {\em LSZ asymptotic condition} holds for the electromagnetic
field).

Huygens' principle was first applied in the context of scattering theory for
the quantized electromagnetic field by Buchholz \cite{Bu1}. More recently, it
was used in \cite{Sp} to prove asymptotic completeness of Rayleigh scattering
in some simple models of electrons permanently confined to nuclei.

Cook's method was used to prove existence of asymptotic fields in simple models
of quantum field theory by Hoegh--Krohn in \cite{H-K}. His arguments were inspired by
work on scattering theory in the context of axiomatic field theory, in
particular Haag--Ruelle scattering theory \cite{Jo} and Hepp's analysis of the LSZ
asymptotic condition in massive quantum field theories \cite{He}. For models of the
kind considered in this paper describing non--relativistic or
pseudo--relativistic electrons interacting with {\em massive} photons, in a
space--time of dimension four or more, {\em Cook's method} (in conjunction with
field--operator domain estimates of the type proven in Lemma
\ref{lemma:estim_higher_ord_nonrel}, Sect. \ref{sec:nonrel}, below) is {\em
all it takes} to construct asymptotic electromagnetic fields; (see e.g.
\cite{Fr}).
The reason is that the {\em amplitude} of a {\em spatially localized} excitation
of the electromagnetic field propagates into the {\em interior} of a forward
light cone if the mass of the photon is {\em strictly positive} and hence
locally decays in time at an integrable rate, $\propto t^{-d/2}$, where $d$ is
the dimension of space. This is {\em not} so if the photon is {\em
massless}. Then if $d$ is {\em odd} Huygens' principle tells us that such
excitations propagate along the {\em boundary} of a forward light cone where
their amplitude decays in time only like $t^{-(d-1)/2}$, which is {\em not} integrable
in dimension $d = 3$ \footnote{ There has been some confusion on this point
in the literature (see e.g. \cite{Fr}) which has been brought to our attention
by Alessandro Pizzo}!

If the propagation velocities of charged particles are smaller than the
propagation velocity of light then every excitation of the electromagnetic field
ends up propagating out of the region where the charged particles are
localized and hence does not interact with them, anymore. This feature, along with
a decay $\propto t^{-(d-1)/2}$ of its amplitude, suffices to rescue Cook's
argument for proving the existence of asymptotic electromagnetic field
operators.

Our discussion makes it clear what our main technical work has to consist in:

We must prove mathematically precise bounds on the velocity of propagation of
electrons ({\em "propagation estimates"}, see Theorem
\ref{theorem:prop_est_nonrel}, Sect. \ref{sec:nonrel}) in physically realistic
situations; and

we have to establish the {\em invariance} of certain domains in
Hilbert space under the time evolution on which propagation estimates hold and
which are contained in the domain of definition of products of electromagnetic
field operators (see Lemma \ref{lemma:estim_higher_ord_nonrel},
Sect. \ref{sec:nonrel}).

The results proven in this paper are an essential ingredient in
developing a mathematically rigorous theory of {\em Rayleigh
scattering} in atomic and molecular physics and in proving that an
isolated neutral atom or molecule prepared in an arbitrary,
possibly highly excited {\em bound state} relaxes to its {\em
ground state}\footnote{See \cite{BFS3, GLL} for results concerning
the existence of ground states in atomic physics.}, as time $t$
tends to infinity, by emitting photons. Relaxation to the ground
state and the related phenomenon of "return to equilibrium" at
positive temperature \cite{JaPi1}, \cite{BFS4} play a key r\^{o}le
in attempts to understand dissipative, irreversible behavior in
the quantum theory of open systems. A rather fundamental
ingredient in proving the relaxation of bound states of neutral
atoms and molecules to their ground states is control over the
nature of the spectrum of the basic Hamilton operator governing
the dynamics of such systems, as achieved in \cite{HuSp},
\cite{BFS3}, \cite{BFSS}, \cite{Skib}. However, spectral results
alone are not sufficient to exhibit relaxation of bound states to
a ground state; they must be supplemented by results on Rayleigh
scattering. Our results in this paper are a step in this
direction; (see also \cite{Sp}). They are, moreover, one among
several key ingredients in developing a scattering theory for
unbound, freely moving charged particles ("infra--particles"
\cite{Schr}) interacting with the quantized radiation field; (see
\cite{Fr}, \cite{FrMStr}, \cite{Bu2} for preliminary results in
this direction).

Next, we introduce the models studied in this paper and summarize our main
results in more precise terms.

We consider a system consisting of
$N=1,2,3,\dots$ quantum--mechanical electrons with non--relativistic kinetimatics
under the influence of an external potential, interacting among themselves via
two--body Coulomb repulsion, and minimally coupled to the soft modes of the
quantized electromagnetic field. Throughout this paper we describe the
transverse degrees of freedom of the electromagnetic field in terms of the
quantized electromagnetic vector potential, $A(x,t)$, in the {\em Coulomb
gauge}, i.e., $(\nabla \cdot A) (x,t)=0$.

Purely for reasons of notational simplicity we neglect electron spin.

The {\em Hamilton operator} generating the time evolution of this system is then
given by
\begin{equation*}
H \equiv H_N^{\rm nr} := \sum_{j=1}^{N}\frac{1}{2m}(p_j+e A(x_j))^2+V_N+W_N+H_f
\end{equation*}
where $V_N$ is a sum of one-body
potentials, $v(x_j)$, describing, for example, the field of static nuclei, and $W_N$ accounts
for the interaction among the particles. Concerning $W_N$, we assume that the
interaction of the $j$-th particle with all the other particles is
repulsive and, of course, that $W_N$ is symmetric with respect to
permutations of the particle coordinates. The operator $H_f$ denotes the field
Hamiltonian (electromagnetic field energy), and $A(x)$ is the quantized UV-cutoff vector
potential at the point $x$. The Hamiltonian $H_N^{nr}$ acts on the Hilbert space
$\mathcal{H} = L^2(\R^{3N})\otimes \mathcal{F} $, where
$\mathcal{F}$ is the bosonic Fock space over $L^2(\R^3;\C^2)$. We can
also treat electrons with spin and work with
$[\sigma\cdot(p+eA(x))]^2=(p+eA(x))^2+e\sigma\cdot B(x)$, instead of $(p+eA(x))^2$.
This only burdens the formalism and is therefore omitted. The quantum statistics
(Pauli principle) obeyed by the electrons turns out not to play any r\^{o}le in
the following.

An important r\^{o}le is played by the threshold
\begin{equation}\label{eq:threshold}
\Sigma = \inf\sigma(H_{N-1}^{\rm nr})+ \liminf_{x\to \infty}v(x).
\end{equation}
This is a lower bound for the total energy when one particle has been put to
infinity. Note that $\Sigma=\infty$ occurs when $v(x)$
is confining. At energies below $\Sigma+mc^2/2$, where $c$ is the speed of
light, the kinetic energy of each particle is less than $mc^2/2$ and hence its
speed is less than $c$. We prove a sharp propagation estimate which confirms this
picture. Under this condition it follows that the interaction of the electrons with photons
escaping to infinity decays integrably fast, as $t\to\infty$. As a result, states with asymptotic incoming and outgoing
photons exist at energies below $\Sigma+mc^2/2$. More precisely, defining $a(h)
= \sum_{\lambda=1,2}\int dk \, a_{\lambda}(k) \overline{h (k,\lambda)}$, and $a^{*} (h) = (a(h))^{*}$, the limits
\begin{equation}\label{asy_operator}
a^{\#}_{+}(h) \varphi = \lim_{t\to\infty} e^{iHt}a^{\#}(h_t)e^{-iHt}\varphi
\end{equation}
exist if $E<\Sigma+mc^2/2$, $\varphi$ is in the range of the spectral
projection $\chi (H \leq E)$, and $h$ belongs to $L^2(\R^3;(1+|k|^{-1})dk)\otimes\C^2$.
The operator $a^{\#}_{+}(h)$ stands for either a creation
operator, $a^{*}_{+}(h)$, or an annihilation operator, $a_{+}(h)$, and $h_t=e^{-i\omega t}h$,
with $\omega(k)=c|k|$. Furthermore, the limit
\begin{equation}\label{products}
a^{\#}_{+}(h_1)\ldots a^{\#}_{+}(h_n)
\varphi=\lim_{t\to\infty}e^{iHt}a^{\#}(h_{1,t})\ldots
a^{\#}(h_{n,t})e^{-iHt}\varphi
\end{equation}
exists if
\begin{equation*}
E+\sum_{i}M_i < \Sigma+\frac{1}{2}mc^2
\end{equation*}
where the sum extends over those $i\geq 2$ only for which $a^{\#}(h_i)$ is a creation
operator. Here $M_i = \sup \{ |k| : h_i (k) \neq 0 \}$. Hence, if
$\Sigma=\infty$ asymptotic creation-- and annihilation
operators are densely defined. They can then be obtained by differentiating
asymptotic Weyl operators. If $\Sigma = \infty$ the operator
$\phi_{+}(h)=1/ \sqrt{2} (a_{+}(h)+a_{+}^{*}(h))$ can be shown to be essentially selfadjoint
on the domain of the Hamilton operator, and
\begin{equation*}
e^{i \phi_{+}(h)}=s-\lim_{t\to\infty}e^{iHt}e^{i \phi (h_t)}e^{-iHt}
\end{equation*}
exists. The asymptotic creation-- and annihilation operators,
$a_{+}^{\#}(h)$, determine a representation of the canonical
commutation relations (CCR) on $\mathcal{H}$. If $\ph_0$ is the
 unique ground state of the Hamilton operator $H$,
i.e., $\ph_0$ is the (up to a phase) unique normalized vector in
$\H$ with $H\ph_0 = E_0 \ph_0$, where $E_0 = \inf \sigma (H)$,
(see \cite{BFS3, GLL, Hiro}), then $a_{+} (h) \ph_{0} = 0$, for an
arbitrary function $h \in L^{2} (\R^{3}, (1+|k|^{-1}) \, dk \, )
\otimes \C^2$. Hence $\ph_0$ is a {\em vacuum} for the asymptotic
creation-- and annihilation operators.

One expects that if $\Sigma = \infty$ {\em asymptotic completeness} (AC) holds,
in the sense that the linear space , $\mathcal{H_{+}}$, spanned by vectors of
the form
\begin{equation}\label{eq:scatt_vectors}
a_{+}^{*} (h_1) \dots a_{+}^{*} (h_n) \ph_0 , \quad n=0,1,2,\dots,
\end{equation}
is {\em dense} in $\mathcal{H}$. A result of
this type, for a simple caricature of the models studied in this paper, has
been proven in \cite{Sp}.

If $\Sigma < \infty$ the situation is more subtle. We define $\mathcal{H}_{bs}$
to be the subspace of $\mathcal{H}$ consisting of "bound states": A vector $\ph
\in \mathcal{H}$ is a {\em bound state} iff it is in the domain of definition of
the operator
\begin{equation*}
\exp \left\{ \eps (\sum_{j=1}^{N} |x_j|) \right\},
\end{equation*}
for some $\eps =\eps(\ph)>0$.
Thus, in a bound state, all electrons are exponentially well localized near the
origin. The space $\mathcal{H}_{+}^{(R)}$ is defined to be the space spanned by
all vectors of the form (\ref{eq:scatt_vectors}), which are strong limits of
vectors of the form
\begin{equation*}
e^{i(H-E_0)t} a^{*} (h_{1,t}) \dots a^{*} (h_{n,t}) \ph_0 ,
\end{equation*}
as $t \to \infty$, where $\ph_0$ is the ground state.

{\em Asymptotic completeness of Rayleigh scattering} is the statement that
\begin{equation}\label{eq:AC_Rayleigh}
\mathcal{H}_{bs} \subseteq \mathcal{H}_{+}^{(R)}.
\end{equation}

We now show that every state of the form $\psi_t = e^{-iHt} \psi$,
with $\psi \in \mathcal{H}_{+}^{(R)}$, relaxes to the ground state
$\ph_0$, as time $t$ tends to $\infty$. If asymptotic completeness
of Rayleigh scattering, Eq. (\ref{eq:AC_Rayleigh}), holds then the
same is true for an {\em arbitrary bound state} $\psi \in
\mathcal{H}_{bs}$.

We must first clarify what is meant by "{\em relaxation of}
$\psi_t$ {\em to the ground state}". Let $\mathcal{A}$ denote the
$C^{*}$ algebra of all bounded functions of the self-adjoint
operators
\begin{equation*}
\phi (h) = \frac{1}{\sqrt{2}} (a(h) + a^{*} (h) ),
\end{equation*}
with $h \in \mathcal{S} (\R^3 ; \C^2 )$, the Schwartz space of two--component
test functions. By taking sums of tensor products of operators in $\mathcal{A}$
with arbitrary bounded operators acting on the Hilbert space of the $N$
electrons one obtains a $C^{*}$ algebra $\tilde{\mathcal{A}}$.

Let $\psi_t = e^{-iHt} \psi$. "Relaxation of $\psi_t$ to the
ground state" is the statement that
\begin{equation}\label{eq:rel_to_gs}
\lim_{t \to \infty} \sprod{\psi_t}{A\psi_t} = \sprod{\ph_0 }{A\ph_0} \,
\sprod{\psi}{\psi} ,
\end{equation}
for an arbitrary operator $A \in \tilde{\mathcal{A}}$.

Let us sketch the proof of (\ref{eq:rel_to_gs}); more details and a proof of
different (generalized) versions of AC of Rayleigh scattering will appear
elsewhere.

A vector $\psi \in \mathcal{H}_{+}^{(R)}$ can be approximated in norm by sums of
vectors of the form
\begin{equation}\label{eq:psin+}
\psi_{n, +} := a_{+}^{*} (h_1) \dots a_{+}^{*} (h_n) \ph_0,
\end{equation}
$h_1, \dots ,h_n $ in $\mathcal{S} (\R^3 ; \C^2 )$, which are strong limits of
the vectors
\begin{equation}\label{eq:str_lim}
e^{i(H-E_0) t} \psi_{n,t},
\end{equation}
where
\begin{equation}\label{eq:psint}
\psi_{n, t} := a^{*} (h_{1,t}) \dots a^{*} (h_{n,t}) \ph_0,
\end{equation}
as $t \to \infty$. Thus, it is enough to show that, for $\psi =\psi_{n,+}$ as
in (\ref{eq:psin+}), Eq. (\ref{eq:rel_to_gs}) holds, with $\ph_0$ as in
(\ref{eq:psin+}).

It follows from (\ref{eq:str_lim}) and (\ref{eq:psint}) that
\begin{equation*}
\lim_{t \to \infty} \| e^{-i(H-E_0)t } \psi_{n,+} - \psi_{n,t} \| = 0 .
\end{equation*}
Thus, for an arbitrary operator $A \in \tilde{\mathcal{A}}$ (which is {\em
bounded}, because $\tilde{\mathcal{A}}$ is a $C^*$ algebra),
\begin{eqnarray}
\lim_{t \to \infty} \sprod{e^{-iHt} \psi_{n,+}}{ A e^{-iHt} \psi_{n,+}} &=&
\lim_{t \to \infty} \sprod{e^{-i(H-E_0)t} \psi_{n,+}}{A
e^{-i(H-E_0)t}\psi_{n,+}} \nonumber \\
&=& \lim_{t \to \infty} \sprod{\psi_{n,t}}{ A \psi_{n,t}} \nonumber \\
&=& \lim_{t \to \infty} \sprod{\prod_{j=1}^{n} a^{*} (h_{j,t}) \ph_0 }{A
\prod_{i=1}^{n} a^{*} (h_{i,t}) \ph_0 }.\label{eq:evol_psin+}
\end{eqnarray}
Next, $A$ is the norm--limit of operators $A_{\alpha} \in \tilde{\mathcal{A}}$, as
$\alpha \to \infty$, with the properties that
\begin{enumerate}
\item[i)] $ A_{\alpha} \prod_{i=1}^{n} a^{*} (h_{i,t}) \ph_0$ is in the domain
of definition of $\prod_{j=n}^{1} a(h_{j,t})$, for arbitrary $t$, and
\item[ii)] $s-\lim_{t \to \infty} \, [ \, \prod_{j=n}^{1} a(h_{j,t}) ,
A_{\alpha}] \, \prod_{i=1}^{n} a^{*} (h_{i,t}) \ph_0 = 0 $.
\end{enumerate}
Property i) follows from easy domain estimates (of the kind of Lemma
\ref{lemma:estim_higher_ord_nonrel},
Sect. \ref{sec:nonrel}), and property ii) follows from the canonical commutation
relations together with
\begin{equation*}
\lim_{t \to \infty} \sum_{\lambda = 1,2} \int dk \, \overline{h_{j} (k , \lambda
)} g(k, \lambda ) e^{i\omega(k) t} = 0,
\end{equation*}
for arbitrary $g \in \mathcal{S} (\R^3 ; \C^2 )$ and $h_j \in \mathcal{S} (\R^3
; \C^2 )$, $j=1, \dots,n$, and from the fact that $\prod_{j=n}^{1} a(h_{j,t})$
commutes with arbitrary bounded operators acting on the Hilbert space of the
electrons.

It then follows from Eq. (\ref{eq:evol_psin+}) and from i) and ii) that
\begin{equation}\label{eq:A*}
\lim_{t \to \infty} \sprod{e^{-iHt}\psi_{n,+}}{ A e^{-iHt} \psi_{n,+}} =
\lim_{t \to \infty} \sprod{A^* \ph_0}{ \prod_{j=n}^{1} a (h_{j,t})
\prod_{i=1}^{n} a^* (h_{i,t}) \ph_0 }.
\end{equation}
Next, one shows that, for arbitrary functions $f_1, \dots , f_m$ in
$\mathcal{S} ( \R^3 ; \C^2 )$, $m=1,2,\dots,$
\begin{equation*}
s-\lim_{t \to \infty } \prod_{l=1}^{m} a (f_{l,t}) \ph_0 = 0 .
\end{equation*}
This implies that
\begin{equation}\label{eq:rel_to_gs_concl}
\begin{split}
s-\lim_{t \to \infty} \prod_{j=n}^{1} a (h_{j,t})
\prod_{i=1}^{n} a^* (h_{i,t}) \ph_0 &= \ph_0 \, \lim_{t \to \infty}
\sprod{\ph_0}{\prod_{j=n}^{1} a (h_{j,t})
\prod_{i=1}^{n} a^* (h_{i,t}) \ph_0 } \\
&= \ph_0 \, \lim_{t \to \infty} \sprod{\prod_{j=1}^{n} a^* (h_{j,t}) \ph_0}
{\prod_{i=1}^{n} a^* (h_{i,t}) \ph_0 } \\
&= \ph_0 \, \sprod{\prod_{j=1}^{n} a_{+}^* (h_{j}) \ph_0}
{\prod_{i=1}^{n} a_{+}^* (h_{i}) \ph_0 } \\
&= \ph_0 \, \sprod{\psi_{n,+} }{\psi_{n,+}}.
\end{split}
\end{equation}
Eqs. (\ref{eq:A*}) and (\ref{eq:rel_to_gs_concl}) show that
\begin{equation*}
\lim_{t \to \infty} \sprod{e^{-iHt}\psi_{n,+}}{ A e^{-iHt} \psi_{n,+}} =
\sprod{\ph_0}{A \ph_0} \sprod{\psi_{n,+}}{\psi_{n,+}},
\end{equation*}
which is what we have claimed we would prove! (More details will appear in a
companion paper).

Next, we describe our model of electrons with relativistic kinematics
interacting with the quantized radiation field ("pseudo--relativistic
electrons") and summarize our main results for this model. For simplicity
(merely of notation!), we consider a one--electron system. The Hamilton operator
is then given by
\begin{equation*}
H^{\rm rel}= \sqrt{(p+eA(x))^2c^2 + m^2c^4} +V + H_f.
\end{equation*}
An attractive feature of this model is the correct relativistic dispersion
relation for the electron, as explained above. For all finite energies, the group velocity of the
electron is
strictly below the velocity of light, and hence the interaction of the electron
with escaping
photons decays, as $t\to\infty$. Again, we will prove a propagation estimate
making this precise. As a consequence, states with incoming and outgoing photons
exist for {\em arbitrary} energies. More precisely,
\begin{equation}\label{eq:claim1_rel}
a^{\#}_{+}(h) \varphi = \lim_{t\to\infty} e^{iHt}a^{\#}(h_t)e^{-iHt}\varphi
\end{equation}
exists, for arbitrary $\varphi\in D((H+i)^{1/2})$, and, for $\varphi\in
D((H+i)^{n/2})$, Eq. \eqref{products} holds, with $H^{\rm nr}$ replaced by $H^{\rm rel}$.

To illustrate the main ideas behind our proofs for existence of
\eqref{asy_operator} and \eqref{eq:claim1_rel} let us consider a massive,
relativistic particle coupled to a quantized scalar field of bosons,
linearly in annihilation and creation operators. Such a system is described by
\begin{equation*}
 H= \sqrt{p^2c^2+m^2c^4} + V+ \phi(G_x) + H_f,
\end{equation*}
acting on $L^2(\R^3)\otimes\F$, where $p=-i\nabla_x$, $x$ is the coordinate of
the particle, $\phi(G_x)= 1 / \sqrt{2} (a(G_x)+a^*(G_x))$, $G_x(k)=e^{-ik\cdot
x}\kappa(k) / \sqrt{|k|}$ and
$\kappa\in C_0^{\infty}(\R^3)$.
Let $h\in C^{\infty}_0(\R^3\backslash\{0\})$ be the wave function of a photon. To prove existence of
$a_{+}^*(h)\ph=\lim_{t\to \infty}e^{iHt}a^{*}(h_t)e^{-iHt}$, we need
to show that the family of vectors
\(t \mapsto \ph(t):=e^{iHt}a^*(h_t)e^{-iHt}\ph\) satisfies the Cauchy
criterion, as $t\to\infty$.
A convenient sufficient condition for this, known as Cook's argument, is
that \(\int_1^{\infty}\| (d/dt) \, \ph(t)\|\,dt<\infty\). From the equation
$a^*(h_t)=\exp(-iH_ft)a^*(h)\exp(iH_ft)$ it is easy to derive that
\begin{equation}\label{phiprime}
\ph'(t) =  i e^{iHt}[\phi(G_x),a^*(h_t)]e^{-iHt}\ph = \frac{i}{\sqrt{2}} e^{iHt}(G_x,h_t)e^{-iHt}\ph,
\end{equation}
where
\begin{equation*}
(G_x,h_t) = \int dk\, e^{i(k\cdot x-\omega(k)t)}\frac{\kappa(k)}{\sqrt{|k|}} h(k),
\end{equation*}
with $\omega(k)=c|k|$. The problem is that $(G_x,h_t) \propto t^{-1}$, for
$|x|\sim ct$, which is {\em not} integrable in $t$. The physical reason for
this problem has been discussed at the beginning of this introduction.
Away from $|x|=ct$, the inner product $(G_x,h_t)$ decays faster than any negative power of
$t$. In fact, by stationary phase arguments, one shows that
\begin{equation*}
\sup_{||x|-t|\geq \eps t}|(G_x,h_t)| \leq C_{n,\eps} t^{-n},
\end{equation*}
while
\begin{equation*}
\sup_{x}|(G_x,h_t)| \leq K/t,
\end{equation*}
for some finite constants $C_{n, \eps}$ and $K$.
From these two estimates and \eqref{phiprime} it follows that $\|\ph'(t)\|$ {\em
is integrable provided that}
\begin{equation}\label{Cook}
\int_1^{\infty}dt\, \frac{1}{t}\|\chi(1-\eps\leq |x/ct|\leq 1+\eps)e^{iHt}\ph\| <\infty.
\end{equation}
It will turn out that it is enough to prove this for $\ph$ in  a suitable dense
subspace, such
as the subspace of states of finite energy which are also in the domain of
$|x|^{1/2}$.

Note that $|(G_x,h_t)| \leq \const\,t^{-3/2}$, which is integrable, if the bosons have a
positive mass, i.e., if $\omega(k)=\sqrt{c^2k^2+M^2c^4}$, with $M>0$. Hence a sharp
propagation estimate, such as the one in Eq. (\ref{Cook}), must
only be proven for massless bosons, as discussed at the beginning of
this introduction.

Note also that \eqref{Cook} is trivial if $\ph$ is
a {\em bound state}, i.e., if $\sup_{x}\||x|^{\alpha}\ph_t(x)\|<\infty$,
for some $\alpha>0$. Then the integrand is of order $t^{-\alpha-1}$, which is
integrable.

To prove \eqref{Cook} in the general case we use that the particle is massive
and hence propagates with a velocity strictly less than $c$, as mentioned above.
As a consequence, $\ph_t(x)$, for $|x/ct|$ near $1$, decays
sufficiently fast for \eqref{Cook} to hold. In fact, we show that
\begin{equation}\label{eq:prop_est}
\int_1^{\infty}\frac{dt}{t^{\mu}}\|\chi(|x/ct|\geq(1-\eps))e^{-iHt}\ph\|^2 \leq \const\|\expect{x}^{1/2}\ph\|^2
\end{equation}
for $\mu>1/2$, $\ph$ with bounded energy distribution and $\eps$ small
enough. From \eqref{eq:prop_est} and the Schwarz inequality the estimate \eqref{Cook} clearly
follows if $\ph\in D(\expect{x}^{1/2})$. Hence $a_{+}^{*}(h)\ph$ exists, for
such $\ph$ and $h\in C^{\infty}_0(\R^3\backslash\{0\})$. For $\ph\in
D((H+i)^{1/2})$ and $h\in L^2((1+|k|^{-1})dk)$, existence then follows by simple approximation
arguments.

The estimate \eqref{eq:prop_est} actually holds also for $\mu=0$ and, at the
expense of higher powers of $\expect{x}^{1/2}$ on the right side, one could
accommodate even positive powers of $t$ in the integral. To keep the proof
short and simple we refrain from discussing these generalizations.\\

Sections \ref{sec:nonrel} and \ref{sec:halbrel} contain our main results on the
non-relativistic and the relativistic model, respectively. The corresponding
propagation estimates are also contained in these sections, but other technical
prerequisites are deferred to various appendices. In
Section \ref{sec:Weyl} we show that asymptotic Weyl operators $W_{+}(h)$ exist and
that they are generated by asymptotic field operators $\phi_{+}(h)$. This is
done for both models simultaneously. Our main results are
Theorems~\ref{theorem:scatt_states_nonrel},
\ref{theorem:many_photon_states_nonrel} in Section \ref{sec:nonrel},
Theorems \ref{theorem:defandprop}, \ref{theorem:many_phot_states_halbrel} in
Section \ref{sec:halbrel}, and Theorem \ref{theorem:weyl} in Section
\ref{sec:Weyl}.

\section{Non--Relativistic QED}\label{sec:nonrel} 

\subsection{Assumptions and Notations}

To describe $N$ non-relativistic particles interacting with the
quantized radiation field we employ the Hamiltonian
\begin{equation}\label{eq:Hnr}
H \equiv H_N :=
\sum_{j=1}^{N}\frac{1}{2}(p_j+A(x_j))^2+(V+W_N)\otimes 1+1\otimes
H_f
\end{equation}
acting on the Hilbert space $\H = L^{2}(\R^{3N})\otimes\F$ where
$\F$ is the bosonic Fock space over $L^2(\R^3;\C^2)$. The units
are chosen in such a way that the mass of the particles, the speed
of light, and Planck's constant are equal to one, and the charge
has been absorbed in the definition of the quantized vector
potential $A(x)$. The scalar potentials $V$ and $W_N$ are
multiplications with real-valued, locally square integrable
functions on $\R^{3N}$, the space of $N$-particle configurations
$(x_1,\ldots,x_N)$. We assume that
\begin{equation*}
V(x_1,\ldots,x_N) = \sum_{j=1}^N v(x_j)
\end{equation*}
and
\begin{equation*}
W_N(x_1,\ldots,x_N) \geq W_{N-1}(x_1,\ldots,\hat{x}_j,\ldots,x_N)
\end{equation*}
for all $j\in\{1,\ldots,N\}$ and $N\geq 2$ ($W_1=0$). Here
$\hat{x}_j$ indicates that the variable $x_j$ is omitted.
Furthermore $W_N$ is symmetric with respect to permutation of the
particle coordinates and
\begin{equation}
V_{-} \leq \eps(-\Delta)+ C_{\eps}\makebox[5em]{\rm for all}\
\eps>0.
\end{equation}
These assumptions are satisfied, for instance, if $v$ is the
Coulomb potential due to static nuclei and $W$ is the Coulomb
repulsion between charged identical particles. The operator $H_f$
measures the field energy and is formally given by
\begin{equation}
H_f = \sum_{\lambda=1,2}\int dk\, |k|
a_{\lambda}^{*}(k)a_{\lambda}(k).
\end{equation}

The interaction between particles and field is described by the
quantized UV-cutoff vector potential in Coulomb gauge
\begin{equation}\label{eq:Apot}
A(x) = \sum_{\lambda=1,2}\int dk\, \frac{\kappa(k)}{\sqrt{2
|k|}}\eps_{\lambda}(k)\left\{e^{ik\cdot x}a_{\lambda}(k)
  + e^{-ik\cdot x}a_{\lambda}^{*}(k)\right\} = \phi(G_x)
\end{equation}
where the polarization vectors $\eps_{\lambda}(k)$ are
perpendicular to $k$ and the form factor $\kappa$ is compactly
supported. In addition we assume  $\kappa\in
C_{0}^{\infty}(\R^3)$. In Eq.~\eqref{eq:Apot},
\(\phi(G_x)=1/\sqrt{2}(a(G_x)+a^{*}(G_x))\) where
\(G_{x}(k,\lambda)=\kappa(k)/\sqrt{|k|}\eps_{\lambda}(k)e^{-ik\cdot
x}\). The creation and annihilation operators $a^*(h)$ and $a(h)$
obey the usual CCR
\begin{equation*}
[a(g),a^*(h)] = (g,h),\hspace{3em}[a^{\sharp}(g),a^{\sharp}(h)]= 0
\end{equation*}
where $(g,h)$ denotes the inner product in $L^2(\R^3;\C^2)$. These
operators are unbounded, they are however bounded w.r.~to
$(H_f+1)^{1/2}$ when $h$ belongs to the weighted $L^2$-space
\begin{align*}
L^2_{\omega}(\R^3;\C^2) &:= \left\{h\in L^2\left|\|h\|_{\omega}^2
:= \sum_{\lambda=1,2}\int dk\,
|h(k,\lambda)|^2(1+|k|^{-1})<\infty\right\}\right.
\end{align*}
In fact, by definition of $a(h)$, $H_f$, by the Schwarz
inequality, and by the CCR
\begin{equation*}
\|a^{\sharp}(h)\ph\| \leq \|h\|_{\omega} \|(H_f+1)^{1/2}\ph\|.
\end{equation*}

The Eq.~\eqref{eq:Hnr} defines the Hamiltonian $H$ as a symmetric
operator on a suitable dense subspace of $\H$. In order to realize
it self-adjointly we need that by Lemma \ref{lemma:eps-bound} (see
Appendix \ref{sec:a_bound})
\begin{equation}\label{eps-bound}
V_{-} \leq \eps H + D_{\eps}\makebox[4em]{for all}\eps>0.
\end{equation}
This implies that $H$ is bounded from below and hence allows us to
define a self-adjoint operator, also called $H$, by the
Friedrichs' extension of $H$. Note that Lemma
\ref{lemma:eps-bound} also implies that all positive parts of $H$
are form bounded with respect to $H$.

\subsection{A Sharp Propagation Estimate}
\label{subsec:prop_est_nonrel}

As explained in the introduction at energies below
$\Sigma+v^2/2=\Sigma+mv^2/2$, $\Sigma$ being the threshold defined
in Eq.~\eqref{eq:threshold}, the kinetic energy of each particle
is less than $v^2/2$ and hence its speed is strictly smaller than
$v$. The purpose of this section is to prove a sharp propagation
estimate which implements this classical argument in our
non-relativistic model of QED. For similar results in $N$-particle
scattering see \cite{SiSo, Sk91, Ge}.

\begin{theorem}\label{theorem:prop_est_nonrel}
Suppose $f \in \mathcal{C}_{0}^{\infty} (\mathbb{R})$, $v>0$, and
$\sup\{\lambda|f(\lambda)\neq 0\} < \Sigma + v^2/2$. Let $\mu >
1/2$ be a fixed constant. Then there exists a constant $C$ such
that
\begin{equation*}
\int_{1}^{\infty} dt \, \frac{1}{t^{\mu}} \, \| \chi (|x_j | \geq
vt) e^{-iHt} f(H) \ph \|^{2} \leq C \, \| (1+|x_j|)^{1/2}f(H)\ph
\|^2
\end{equation*}
for all $j\in\{1,\ldots,N\}$ and all $\ph \in \mathcal{H}$.
\end{theorem}
{\em Remarks.}  1) We are most interested in the case $v=1-\eps$
where $\eps>0$ but may be chosen as small as we please. Then it
suffices to assume $\supp(f)\subset(-\infty , \Sigma + 1/2)$.

2) The theorem actually holds for $\mu = 0$, but, for $\mu > 1/2$,
the proof is easier, and this result is sufficient for our
purposes.

\begin{proof}
We assume $f$ is real-valued; the proof for complex-valued $f$ is
similar. Let $\eps>0$ with $\eps\leq v$. Further restrictions will
be imposed later. Pick $h \in \mathcal{C}^{\infty} (\R)$,
non-decreasing, with $0 \leq h \leq 1$, $h(s)=1$ if $s \geq v$,
and $h(s)=0$ if $s\leq v-\eps$. The function $\tilde{h} (s) =
\int_{0}^{s} d\tau h^2 (\tau )$ is a smooth version of
$(s-v)\chi(s-v\geq 0)$. By construction of $h$,
\begin{equation}\label{eq:htildeh}
\tilde{h}(s) \leq (s-(v-\eps))h^2(s).
\end{equation}
Henceforth $h$ and $\tilde{h}$ are abbreviations for
$h(\expect{x_j}/t )$ and $\tilde{h}(\expect{x_j}/t)$ respectively
and $\expect{x_j}= (1+x_j^2)^{1/2}$. We will also use the notation
$\ph_t = e^{-iHt} \ph$. The operator
\begin{equation*}
\phi(t) = - f (H)\,  t^{1-\mu} \tilde{h}\, f(H)
\end{equation*}
will play the role of the so called propagation observable. We
will show that its Heisenberg derivative
\(D\phi(t)=[iH,\phi(t)]+\partial/(\partial t)\phi(t)\) satisfies
\begin{equation}\label{eq:nrtodo}
D\phi(t) \geq \delta \, t^{-\mu} \, f(H)h^2f(H) +
(\text{integrable w.r.t. $t$})f(H)^2 ,
\end{equation}
for $t \in [T_0 , \infty )$, where $\delta>0$ and $T_0 > 1$ is
sufficiently large. It will follow for $T \geq T_0$ that
\begin{equation}\label{eq:prop_obs}
\begin{split}
\int_{T_0}^{T}dt \, &t^{-\mu} \, \sprod{\ph_t}{f (H)h^2f(H)\ph_t}
\\ &\leq \frac{1}{\delta}\left\{ \sprod{\ph_T}{\phi (T) \ph_T}
-\sprod{\ph_{T_0}}{\phi (T_0) \ph_{T_0}} \right\} + C
\|f(H)\ph\|^2
\\ &\leq \frac{1}{\delta} |\sprod{\ph_{T_0}}{\phi (T_0)
\ph_{T_0}}| + C \| f(H)\ph \|^{2},
\end{split}
\end{equation}
where, in the last step, we used that $\sprod{\ph_T}{\phi (T)
\ph_T} \leq 0$, for all $T>1$. This estimate proves the theorem
because $h \geq \chi(|x_j|\geq vt)$ and because
$|\sprod{\ph_{T_0}}{\phi(T_0)\ph_{T_0}}|\leq \const
\|(1+|x_j|)^{1/2}f(H)\ph \|^2$ by Lemma~\ref{lemma:invar} applied
to $\exp(-iHT_0)g(H)$ with $g\in C_0^{\infty}(\R)$ such that
$g(H)f(H)=f(H)$. In order to prove (\ref{eq:nrtodo}) we note that
$D\phi(t) = -f(H)\{[iH,t^{1-\mu} \tilde{h}]+ (
\partial /
\partial t)(t^{1-\mu} \tilde{h})\}f(H)$, where by construction of
$\tilde{h}$ and by \eqref{eq:htildeh}
\begin{equation}\label{eq:partial_tildeh}
-\frac{\partial}{\partial t}(t^{1-\mu} \tilde{h}) \geq (v-\eps )
\, t^{-\mu} \, h^2
\end{equation}
and, with the notation $\pi(x_j)=p_j+A(x_j)$,
\begin{align}
[iH,t^{1-\mu}\tilde{h}] &=
\frac{t^{1-\mu}}{2}\left\{\pi(x_j)\cdot\nabla\tilde{h} +
  \nabla\tilde{h}\cdot \pi(x_j)\right\}\nonumber  \\
 &= \frac{t^{-\mu}}{2} h\left\{\pi(x_j)\cdot\frac{x_j}{\expect{x_j}} +
  \frac{x_j}{\expect{x_j}}\cdot \pi(x_j)\right\}h.\label{eq:Hcommut}
\end{align}
In the last equation we used that $\nabla\tilde{h}=h^2\,
x_j/\expect{x_j} \, t^{-1}$ and then commuted one factor $h$ to
the left and the second one to the right. The commutators which
arise cancel. By (\ref{eq:Hcommut})
\begin{equation}\label{eq:nrpe1}
|\sprod{\ph_t}{f(H) [iH,t^{1-\mu}\tilde{h}]f(H)\ph_t}| \leq
t^{-\mu} \, \| h f(H)\ph_t\| \,  \|\pi (x_j) h f(H) \ph_t\| ,
\end{equation}
where we used Schwarz's inequality and $| x_j  / \expect{x_j} |
\leq 1$. Consider the factor $\|\pi (x_j) h f(H) \ph_t\|$ on the
r.h.s. of the last equation. From $|x_j|\geq (v-\eps)t$ on the
support of  $h$, the definition of $\Sigma$, and $W_N-W_{N-1}\geq
0$ it is easy to see that
\begin{align}\label{eq:nrpe2}
h\pi(x_j)^2h &\leq
     2h\left\{\frac{1}{2}\pi(x_j)^2+v(x_j)+W_N-W_{N-1}+H_{N-1}-\Sigma+o(t^{0})\right\}h
\nonumber \\ &= 2h\left\{H-\Sigma+o(t^0)\right\} h.
\end{align}
Next pick $g=\bar{g}\in C_0^{\infty}(\R)$ with $gf=f$ and
\(\supp(g)\subset (-\infty,E+\eps/2]\) where $E=\sup \{ \lambda :
f(\lambda )\neq 0 \}$, so that $f(H)=f(H)g(H)$ and $g(H)Hg(H)\leq
g(H)(E+\eps/2)g(H)$. Then $[g(H),h] = O(t^{-1})$ implies that
\begin{equation}\label{eq:HtoE}
f(H) h H h f(H) \leq f(H) h (E+\eps/2) h f(H) + O (t^{-1})f(H)^2.
\end{equation}
Equations \eqref{eq:nrpe2} and \eqref{eq:HtoE} combined show that
\begin{equation}\label{eq:velo_energy}
\|\pi (x_j) h f(H) \ph_t\| \leq \; (2(E-\Sigma+\eps))^{1/2} \,
\|hf(H)\ph_t\| + C \,  t^{-1/2} \, \| f(H)\ph \| ,
\end{equation}
for $t$ large enough. Inserting this into \eqref{eq:nrpe1} we get
\begin{equation}\label{eq:iH_tildeh}
\begin{split}
| \sprod{\ph_t}{f(H)[iH,\tilde{h}]f(H)\ph_t} | \leq \; &t^{-\mu}
\, (2(E-\Sigma+\eps))^{1/2} \, \|hf(H)\ph_t\|^2 \\ &+ C\, t^{-1/2
- \mu} \, \| f(H)\ph \|^2 ,
\end{split}
\end{equation}
which, together with (\ref{eq:partial_tildeh}), shows that
\begin{equation}\label{eq:last}
\begin{split}
\sprod{\ph_t}{D\phi (t) \ph_t} \geq \; t^{-\mu} \, &\{ (v -\eps )
- \sqrt{2(E - \Sigma + \eps )} \} \, \| h f(H) \ph_t \|^2 \\ &- C
\, t^{-\mu - 1/2} \, \| f(H)\ph \|^{2}.
\end{split}
\end{equation}
We choose now $\eps > 0$ so small that $ E+\eps < \Sigma +
(v-\eps)^2 /2$. Then (\ref{eq:nrtodo}) follows from
(\ref{eq:last}) with $\delta =  \{ (v -\eps ) - \sqrt{2(E - \Sigma
+ \eps )} \} > 0$.
\end{proof}

\subsection{Existence of Asymptotic Field Operators}
\label{subsec:existence_nonrel}

Next we use the propagation estimate from
Theorem~\ref{theorem:prop_est_nonrel} to establish existence of
the asymptotic field operators $a_{+}^*(h)$ and to show that
scattering states with an arbitrary number of asymptotically free
photons exist. The main results are summarized in the
Theorems~\ref{theorem:scatt_states_nonrel} and
\ref{theorem:many_photon_states_nonrel}. An important and
non-trivial technical ingredient for the scattering of $n$ photons
is the boundedness of $H_f^n(H+i)^{-n}$ which is proved in
Appendix \ref{sec:est_high_ord_nonrel}.

In addition to Theorem~\ref{theorem:prop_est_nonrel} we need the
following lemma (see \cite[Theorem XI.18]{RSIII}).

\begin{lemma}\label{lemma:phot_disp}
Suppose $h \in \mathcal{C}_{0}^{\infty} ( \R^{3} \backslash \{ 0
\} )$. Then there exists a constant $C$ such that
\begin{equation}\label{eq:beh1}
\sup_{ x \in \R^{3}} \left| \int dk  \, h (k) \, e^{i( k \cdot x -
| k | t)} \right| \leq  \, \frac{C}{| t |},
\end{equation}
for all $ t \in \R$.
\end{lemma}

\begin{proposition}\label{proposition:exist_nonrel}
Suppose $f \in \mathcal{C}_{0}^{\infty} (\mathbb{R})$, with $
\supp (f ) \subset ( -\infty \, , \, \Sigma + 1/2 )$ and $h \in
\mathcal{C}_{0}^{\infty} (\mathbb{R}^{3} \backslash \{ 0 \} ;
\mathbb{C}^2 )$. Then, for all $\ph \in \mathcal{H}$, the limit
\begin{equation}\label{eq:to_prove}
\lim_{t \to \infty} e^{iHt} a^{\sharp} (h_{t}) e^{-iHt} f(H) \ph
\end{equation}
exists.
\end{proposition}
\begin{proof}
Since $e^{iHt} a^{\sharp} (h_t) e^{-iHt} f(H)$ is bounded
uniformly in $t$ it suffices to prove existence of
(\ref{eq:to_prove}) for vectors $\ph$ from the dense subspace
$D(\expect{x}^{1/2})$. We only consider creation operators, the
proof for annihilation operators is similar. For given $\ph \in
\mathcal{H}$ let
\begin{equation*}
\ph (t) = e^{iHt} \, a^{*} (h_{t}) \, f \, e^{-iHt} \ph ,
\end{equation*}
where $f = f(H)$. By Cook's argument the existence of the limit
(\ref{eq:to_prove}) follows if
\begin{equation*}
\int_{1}^{\infty} dt \, \| \frac{d}{dt} \ph (t) \| < \infty .
\end{equation*}
In the following we will use the notation $\ph_t = e^{-iHt} \ph$.
A straightforward computation shows that
\begin{equation}\label{eq:d_dt_ph_t}
\begin{split}
\frac{d}{dt} \, \ph (t) &= \frac{1}{2}\sum_{j=1}^{N} \, e^{iHt} \,
 [i (p_{j}+ A (x_{j}))^{2} , a^{*} (h_{t})] \, f \ph_{t} \\
&= \frac{i}{\sqrt{2}} \sum_{j=1}^{N} e^{iHt} \, (G_{x_{j}} ,
h_{t}) \cdot  (p_{j}+ A (x_{j})) \, f \ph_{t}.
\end{split}
\end{equation}
Next we fix $j \in \{ 1, \dots N \}$ and show that the
corresponding term has norm which is integrable w.r.t. $t$. Choose
$\eps > 0$ and so small that $\sup \{ \lambda : f (\lambda ) \neq
0 \} < \Sigma + (1-2\eps)^2 /2$. Pick then $ \chi_{1}$, $\chi_{2}
\in \mathcal{C}^{\infty} (\mathbb{R})$, $ 0 \leq \chi_{i} \leq 1$,
$ \chi_{1}^2 + \chi_{2}^{2} =1$, $\chi_{1} (s) =0$ if $ s \leq 1-
2 \eps $ and $\chi_{1} (s) =1$ for $ s \geq 1-\eps$. Let
$\chi_{1}=\chi_{1} ( |x_j | / t )$ and $\chi_{2} = \chi_{2} ( |x_j
| / t )$ henceforth. The term in the sum on the r.h.s. of
(\ref{eq:d_dt_ph_t}) corresponding to the fixed $j$ then has norm
bounded by
\begin{equation}\label{eq:k12}
\| e^{iHt} \, (G_{x_{j}} , h_{t}) \cdot  (p_{j}+ A (x_{j})) \, f
\ph_{t} \| \leq \frac{1}{\sqrt{2}} \, \sum_{k=1}^{2} \,  \|
(G_{x_{j}} ,h_{t}) \chi_{k} \| \, \| \chi_k (p_j + A(x_j))
 f \ph_{t} \| .
\end{equation}
Consider first the $k=2$ term. Since $\| (p_{j} + A (x_{j}) ) f
(H) \| < \infty$, and since, by a stationary phase argument,
$\sup_{x_{j}} | ( G_{x_{j}} , h_t ) \chi_{2} | \leq C_{n} / t^n$,
for any $n \geq 1$, the term with $k=2$ on the r.h.s. of
(\ref{eq:k12}) is integrable w.r.t. $t$. Consider now the term
with $k=1$. We first note, that
\begin{equation}\label{eq:comm_chi_p}
\| \chi_1 \, (p_j + A (x_j)) \, f \, \ph_t \| \leq  \| (p_j + A
(x_j)) (H+i)^{-1} \| \, \| \chi_1 \tilde{f} \ph_t \| + C / t ,
\end{equation}
where $\tilde{f} = (H+i) f$. In order to prove the last equation,
write $f=(H+i)^{-1} \tilde{f}$ and then commute the operator $(p_j
+ A(x_j)) (H+i)^{-1}$ to the left of $\chi_1$. The factor
proportional to $t^{-1}$ on the r.h.s. of (\ref{eq:comm_chi_p})
arises from the commutator of these two operators. Since, by Lemma
\ref{lemma:phot_disp}, $\| \chi_1 (G_{x_j} , h_t )\| \leq C / t$,
it follows that the term with $k=1$ on the r.h.s. of
(\ref{eq:k12}) is bounded by $\const \, t^{-1}  \, \| \chi_1
\tilde{f} \ph_t \| + \const \, t^{-2}$. The term proportional to
$t^{-2}$ is clearly integrable w.r.t. $t$ and by the Schwarz
inequality
\begin{equation}
\int_{1}^{\infty} dt \, \frac{1}{t} \, \| \chi_1 \tilde{f} \ph_t
\| \leq \left( \int_1^{\infty} dt \, \frac{1}{t^{1+2\delta}}
\right)^{\frac{1}{2}} \left( \int_1^{\infty} dt \,
\frac{1}{t^{1-2\delta}} \, \| \chi_1 \tilde{f} \ph_t \|^2
\right)^{\frac{1}{2}}.
\end{equation}
If we choose $\delta \in (0 , 1/4)$ this is finite by Theorem
\ref{theorem:prop_est_nonrel} with $\mu = 1-2\delta > 1/2 $ and
$v=1-2\eps$, because $\tilde{f} (H) \ph \in D(\expect{x}^{1/2})$
by Lemma \ref{lemma:invar}.
\end{proof}

In the following theorem we list some important properties of the
asymptotic field operators.

\begin{theorem}\label{theorem:scatt_states_nonrel}
Suppose that $\ph = \chi (H \leq E)\ph$, for some $E < \Sigma +
1/2$, and that $h,g \in L_{\omega}^{2} (\mathbb{R}^{3};
\mathbb{C}^2 )$.
\begin{itemize}
\item[i)]The limit
\begin{equation*}
a_{+}^{\sharp} (h) \ph = \lim_{t \to \infty} e^{iHt} a^{\sharp}
(h_{t}) e^{-iHt}\ph
\end{equation*}
exists, and furthermore
\begin{equation}\label{eq:a_H_bound}
\| \, a_{+}^{\sharp} (h) \chi (H \leq E) \| \leq C \, \| h
\|_{\omega},
\end{equation}
for some finite constant $C > 0$.
\item[ii)] The canonical commutation relations
\begin{equation*}
[ a_{+} (g) , a_{+}^{*} (h) ] = (g,h) \qquad \text{and} \qquad
[a_{+}^{\sharp}(h) , a_{+}^{\sharp}(g)] = 0 ,
\end{equation*}
hold true, in form--sense, on $\chi (H \leq E) \mathcal{H}$.
\item[iii)]If in addition $\omega h \in L^{2}
(\mathbb{R}^{3}; \C^2 )$, then
\begin{equation*}
[H , a_{+}^{*} (h) ] = a_{+}^{*} (\omega h), \qquad  [ H, a_{+}
(h) ] = -a_{+} (\omega h),
\end{equation*}
in form--sense on $\chi (H\leq E) \mathcal{H}$. Moreover if
$\Sigma = \infty$ then the operator
\begin{equation*}
\phi_{+} (h) = \frac{1}{\sqrt{2}} \, ( a_{+}(h) + a_{+}^* (h) )
\end{equation*}
is essentially self--adjoint on $\cup_{d  > 0} \chi (H \leq d )
\mathcal{H}$, and the domain of its closure contains
$D((H+i)^{1/2})$.
\item[iv)]Let $m=\inf \{ | k| : h(k) \neq 0 \}$ and $M=\sup \{ |k| : h(k) \neq 0
\}$. Then
\begin{equation*}
\begin{split}
a_{+}^{*} (h) \Ran \chi (H \leq E) &\subset \Ran \chi (H\leq E+M)
\\ a_{+} (h) \Ran \chi (H \leq E) &\subset \Ran \chi (H\leq E-m )
.
\end{split}
\end{equation*}
\end{itemize}
\end{theorem}
\begin{proof}
\begin{itemize}
\item[i)] The existence of the limit for $h \in \mathcal{C}_{0}^{\infty} (\R^{3}
\backslash \{ 0 \}; \C^2 )$ follows from Proposition
\ref{proposition:exist_nonrel}. The operator bound
(\ref{eq:a_H_bound}) follows from $\| \, a^{\sharp} (h_{t}) (H_{f}
+ 1)^{-1/2} \| \leq \| \, h \|_{\omega}$ and from the boundedness
of $(H_{f} + 1)^{1/2} (H+i)^{-1/2}$. Finally if $h \in
L_{\omega}^{2}(\R^3 ; \C^2 )$ the existence of the limit follows
by an approximation argument, because $\mathcal{C}_{0}^{\infty}
(\mathbb{R}^{3} \backslash \{ 0 \} )$ is dense in $L_{\omega}^{2}
(\R^{3})$.
\item[ii)] Follows from i) and the CCR for $a(h)$ and $a^{*}
(h)$.
\item[iii)] For $\ph$ and $\psi \in \chi (H \leq E ) \mathcal{H}$
\begin{equation*}
\begin{split}
\sprod{\ph}{[H,a_{+}^{*} (h)] \psi} &= \lim_{t \to \infty}
\sprod{\ph_{t}}{ [H,a^{*} (h_{t})] \psi_{t}} \\ &= \lim_{t \to
\infty} \frac{1}{2} \sum_{j=1}^{N} \, \sprod{\ph_{t}}{ [ (p_{j} +
A (x_{j}))^{2} , a^{*} (h_{t})] \psi_{t}} + \sprod{\ph}{a_{+}^{*}
(\omega h) \psi} \\ &= \sprod{\ph}{a_{+}^{*} (\omega h) \psi},
\end{split}
\end{equation*}
because $|\sprod{\ph_{t}}{[ (p_{j} + A (x_{j}))^{2} , a^{*}
(h_{t})] \psi_{t}}| \leq  C \, \sup_{x_{j}} | (G_{x_{j}} , h_{t})|
\to 0$ as $t \to \infty$. The proof of the second pull--through
formula is similar. The proof of the essentially self--adjointness
of $\phi_{+} (h)$, in the case $\Sigma = \infty$, is very similar
to the proof of the corresponding statement for the
pseudo--relativistic model (see Theorem \ref{theorem:defandprop},
iii)).
\item[iv)] From the spectral theorem we know that $\psi \in \Ran \chi ( H \leq
\lambda )$ if and only if the function $\R \ni t \to e^{iHt} \psi$
has an analytic extension $z \to e^{iHz} \psi$, for $\Ima z < 0$,
which satisfies $\| e^{iHz} \psi \| \leq C e^{| \Ima z |
\lambda}$. Now pick $\psi \in \Ran \chi (H \leq E)$. Then we have,
from iii),
\begin{equation*}
e^{iHt} a_{+}^{*} (h) \psi = a_{+}^{*} ( e^{i\omega t } h) e^{iHt}
\psi.
\end{equation*}
Since $H$ and $\omega$ are bounded from below, it follows that the
function $\R \ni t \to e^{iHt} a_{+}^{*} (h) \psi$ has an analytic
extension, given by, $a_{+}^{*} (e^{i\omega z} h) e^{iHz} \psi$,
which satisfies
\begin{equation*}
\begin{split}
\| a_{+}^{*} (e^{i\omega z} h) e^{iHz} \psi \| &\leq \| a_{+}^{*}
(e^{i\omega z} h) \chi (H \leq E) \| \, \| e^{iHz} \psi \| \\
&\leq C \|e^{i\omega z} h \|_{\omega}  e^{| \Ima z | E } \\ &\leq
C e^{|\Ima z| (E+M)},
\end{split}
\end{equation*}
where we used the estimate (\ref{eq:a_H_bound}). This proves that
$a_{+}^{*} (h) \psi \in \Ran \chi (H \leq E+M)$. The statement for
annihilation operators can be proved similarly.
\end{itemize}
\end{proof}

Existence of scattering states with an arbitrary number of
escaping photons does not obviously follow from
Theorem~\ref{theorem:scatt_states_nonrel} because creation and
annihilation operators are unbounded. They are however bounded
with respect to $(H_f+1)^{1/2}$ and by the following lemma powers
of $H_f$ are bounded w.r.~to powers of $H$.

\begin{lemma}\label{lemma:estim_higher_ord_nonrel}
For any integer $n \geq 1$, the operators
\begin{itemize}
\item[i)] $[H_{f}^{n-1} , H ] (H+i)^{-n}$
\item[ii)] $H_{f}^{n} (H+i)^{-n}$
\end{itemize}
are bounded.
\end{lemma}

{\em Remark.} By an abstract interpolation argument it follows
that also $H_{f}^{n/2} (H+i)^{-n/2}$ is bounded for all
non-negative integers.

This lemma, whose proof is deferred to
Appendix~\ref{sec:est_high_ord_nonrel}, is the main technical
ingredient, apart from Theorem~\ref{theorem:scatt_states_nonrel},
for the proof of the following theorem.

\begin{theorem}\label{theorem:many_photon_states_nonrel}
Suppose $\ph = \chi (H \leq E ) \ph$, $h_{i} \in L_{\omega}^{2}
(\R^{3}; \C^2 )$ and let $M_{i} = \sup \{ |k| : h_{i} (k) \neq 0
\}$, for $i \in \{ 1,\dots ,n \}$. Then
\begin{equation}\label{eq:multi_phot_lim}
a_{+}^{\sharp} (h_{1} ) \dots a_{+}^{\sharp} (h_{n}) \ph = \lim_{t
\to \infty} e^{iHt} a^{\sharp} (h_{1,t}) \dots a^{\sharp}
(h_{n,t}) e^{-iHt} \ph ,
\end{equation}
provided that $E + \sum_{i \geq 2} M_{i} < \Sigma + 1/2$, where
the sum $\sum_{i \geq 2} M_{i}$ runs over those $i's$ bigger than
$1$ for which $a^{\sharp} (h_{i})$ is a creation operator.
Furthermore there exists a constant $C_n$, independent of $E$ such
that
\begin{equation}\label{eq:n-fields-bound}
\|a_{+}^{\sharp} (h_{1}) \dots a_{+}^{\sharp} (h_{n})
(H+i)^{-n/2}\chi(H \leq E) \| \leq C_n
\|h_1\|_{\omega}\ldots\|h_n\|_{\omega}.
\end{equation}
\end{theorem}

\begin{proof}
By Theorem \ref{theorem:scatt_states_nonrel}, iv) and because of
the assumption $E + \sum_{i} M_{i} < \Sigma + 1/2$, it follows
that, for each $l \in \{2,\dots ,n \}$ the vector $a_{+}^{\sharp}
(h_{l}) \dots a_{+}^{\sharp} (h_{n}) \ph$ is in the range of a
spectral projector $\chi (H \leq E^{\prime})$, for some
$E^{\prime} < \Sigma +1/2$. Thus $a_{+}^{\sharp} (h_{l-1} ) \dots
a_{+}^{\sharp} (h_{n}) \ph$ is well defined and
\begin{equation}\label{eq:a_+l_n}
a_{+}^{\sharp} (h_{l} ) \dots a_{+}^{\sharp} (h_{n}) \ph = \lim_{t
\to \infty} e^{iHt} a^{\sharp} (h_{l,t}) e^{-iHt} a_{+}^{\sharp}
(h_{l+1} ) \dots a_{+}^{\sharp} (h_{n}) \ph .
\end{equation}
Now we want to prove the equality (\ref{eq:multi_phot_lim}). We
consider only the case with $n$ creation operators, the other
cases being similar. Assume first that $h_{i} \in
\mathcal{C}_{0}^{\infty} (\R^3 \backslash \{ 0 \} ; \C^2 )$, for
all $i \in \{ 1 , \dots , n \}$. We proceed then by induction over
$n$. The statement for $n=1$ follows from Theorem
\ref{theorem:scatt_states_nonrel}. Now we assume that the
statement holds for any integer less than a given $n$ and we prove
that it holds also for the product of $n$ creation operators. To
this end we write
\begin{equation}\label{eq:trenn}
\begin{split}
e^{iHt} a^{*} (h_{1,t}) \dots &a^{*} (h_{n,t}) e^{-iHt}\ph -
a_{+}^{*} (h_{1} ) \dots a_{+}^{*} (h_{n}) \ph \\ = \; &(e^{iHt}
a^{*} (h_{1,t})e^{-iHt} - a_{+}^{*} (h_1)) a_{+}^{*} (h_2) \dots
a_{+}^{*}(h_n) \ph \\ &+ e^{iHt} a^{*} (h_{1,t})e^{-iHt} \times \\
&\left\{ e^{iHt} a^{*} (h_{2,t}) \dots a^{*} (h_{n,t}) e^{-iHt} -
a_{+}^{*} (h_2) \dots a_{+}^{*} (h_n) \right\} \ph .
\end{split}
\end{equation}
The norm of the first term on the r.h.s of the last equation
converges, by (\ref{eq:a_+l_n}), to $0$ as $t \to \infty$. To
handle the second term on the r.h.s. of (\ref{eq:trenn}) we insert
the operator $ id = (H+i)^{-1} (H+i) $ just in front of the
braces, and we commute the factor $(H+i)$ through the operators
within the braces. The second term on the r.h.s. of
(\ref{eq:trenn}) becomes then
\begin{equation}\label{eq:sec_term}
\begin{split}
e^{iHt} &a^{*} (h_{1,t})e^{-iHt} (H+i)^{-1} \times \\ \left\{
\right. &(e^{iHt} a^{*} (h_{2,t}) \dots a^{*} (h_{n,t}) e^{-iHt} -
a_{+}^{*} (h_2) \dots a_{+}^{*} (h_n))(H+i)\ph  \\ &+
\sum_{l=2}^{n} (e^{iHt} a^{*} (h_{2,t}) \dots a^{*} (\omega
h_{l,t} ) \dots a^{*} (h_{n,t}) e^{-iHt} - a_{+}^{*} (h_2) \dots
a_{+}^{*} (\omega h_l) \dots a_{+}^{*} (h_n)) \ph \\
 &+ \sum_{l=2}^{n} \sum_{j=1}^{N}  e^{iHt} a^{*} (h_{2,t}) \dots
\frac{1}{2} [(p_j +A(x_j))^{2} , a^{*} (h_{l,t})] \left. \dots
a^{*} (h_{n,t}) e^{-iHt} \ph \right\} .
\end{split}
\end{equation}
Now the term in front of the braces is bounded uniformly in $t$.
The first term inside the braces, and each factor in the first sum
converge to $0$, as $t \to \infty$, by induction hypothesis.
Finally the terms in the last sum inside the braces have norm
which is bounded by
\begin{equation}\label{eq:norm_bou}
\frac{1}{\sqrt{2}}  \| ( G_{x_{j}} , h_{l,t}) \| \, \| a^{*}
(h_{2,t}) \dots a^* (h_{l-1,t}) (p_j + A(x_j)) a^* (h_{l+1,t})
\dots a^{*} (h_{n,t}) (H + i)^{-n} \| \, \| (H+i)^{n} \ph \|.
\end{equation}
The first factor in (\ref{eq:norm_bou}) converges to 0, as $t \to
\infty$. By Lemma \ref{lemma:estim_higher_ord_nonrel} it follows
that the second factor in (\ref{eq:norm_bou}) is bounded uniformly
in $t$. We have thus shown that both terms on the r.h.s. of
(\ref{eq:trenn}) converge to 0, as $t \to \infty$. This proves
equation \eqref{eq:multi_phot_lim} if $h_i \in
\mathcal{C}_{0}^{\infty} (\R^3 \backslash \{ 0 \} ; \C^2 )$, for
all $i \in \{ 1, \dots, n \}$. For $h_{i} \in L_{\omega}^{2} (\R^3
; \C^2 )$ equation (\ref{eq:multi_phot_lim}) follows now by an
approximation argument, because $\mathcal{C}_{0}^{\infty} (\R^3
\backslash \{ 0 \})$ is dense in $L_{\omega}^{2} (\R^3 )$, by
Lemma \ref{lemma:ann_cre_bound} and because
\begin{equation*}
\| a_{+}^{*} (h_1) \dots a_{+}^{*} (h_n) \chi (H\leq E) \| \leq C
\|h_1 \|_{\omega} \dots \|h_{n} \|_{\omega},
\end{equation*}
provided $E+ \sum_{i \geq 2} M_i < \Sigma + 1/2$. The last
estimate follows from the bound (\ref{eq:a_H_bound}) and from
Theorem \ref{theorem:scatt_states_nonrel} iv), which implies
\begin{equation*}
\begin{split}
a_{+}^{*} (h_1) \dots a_{+}^{*} (h_n) \chi (H\leq E) &= a_{+}^{*}
(h_1)\chi (H \leq E+ \Sigma_{l \geq 2} M_l) \dots \\ &\dots
a_{+}^{*} (h_{n-1}) \chi (H \leq E+M_n ) a_{+}^{*} (h_n)
\chi(H\leq E).
\end{split}
\end{equation*}
Finally \eqref{eq:n-fields-bound} follows from
\eqref{eq:multi_phot_lim}, Lemma~\ref{lemma:ann_cre_bound}, and
Lemma~\ref{lemma:estim_higher_ord_nonrel}. This completes the
proof of the theorem.
\end{proof}

In the case $\Sigma=\infty$ the results of
Theorem~\ref{theorem:many_photon_states_nonrel} can be refined to
get following theorem.

\begin{theorem}\label{theorem:confined}
Suppose $\Sigma=\infty$, $h_i\in L^2_{\omega} (\R^3 ; \C^2 )$ for
$i=1,\ldots,n$ and $\ph\in D((H+i)^{n/2})$. Let
$a_{+}^{\sharp}(h_i)$ denote the closure of the asymptotic
operators $a_{+}^{\sharp}(h_i)$ obtained in
Theorem~\ref{theorem:scatt_states_nonrel} and defined on
$\cup_{d>0}\chi(H\leq d)\H$. Then \(\ph\in D(a_{+}^{\sharp} (h_{1}
) \dots a_{+}^{\sharp} (h_{n}))\),
\[\|a_{+}^{\sharp} (h_{1}) \dots a_{+}^{\sharp} (h_{n})(H+i)^{-n/2}\| \leq C_n
\|h_1\|_{\omega}\ldots\|h_n\|_{\omega}\] and
\begin{equation*}
a_{+}^{\sharp} (h_{1} ) \dots a_{+}^{\sharp} (h_{n})\ph = \lim_{t
\to \infty} e^{iHt} a^{\sharp} (h_{1,t}) \dots a^{\sharp}
(h_{n,t}) e^{-iHt} \ph.
\end{equation*}

\begin{proof}
We proceed by induction over $n$ and assume the statement holds
for $n$ replaced by $n-1$. Let $\ph_d=\chi(H\leq d)\ph$ where
$d\in \R$.Then $(H+i)^{n/2}(\ph_d-\ph)\to\ 0$ as $d\to\infty$.
Hence by Theorem~\ref{theorem:many_photon_states_nonrel} the limit
\begin{equation*}
\lim_{d\to\infty} a_{+}^{\sharp} (h_{1} ) \dots a_{+}^{\sharp}
(h_{n})\ph_d
\end{equation*}
exists and by the induction hypothesis
\begin{equation*}
\lim_{d\to\infty} a_{+}^{\sharp} (h_{2} ) \dots a_{+}^{\sharp}
(h_{n})\ph_d = a_{+}^{\sharp} (h_{2} ) \dots a_{+}^{\sharp}
(h_{n})\ph.
\end{equation*}
Since $a_{+}^{\sharp}(h_1)$ is closed this proves \(a_{+}^{\sharp}
(h_{2} )\dots a_{+}^{\sharp} (h_{n})\in D(a_{+}^{\sharp}(h_1))\)
and the first equation from
\begin{align*}
a_{+}^{\sharp} (h_{1} ) \dots a_{+}^{\sharp} (h_{n})\ph =&
\lim_{d\to\infty}a_{+}^{\sharp} (h_{1} ) \dots a_{+}^{\sharp}
(h_{n})\ph_d\\ =& \lim_{d\to\infty}\lim_{t\to\infty}e^{iHt}
a^{\sharp} (h_{1,t} ) \dots a^{\sharp} (h_{n,t})e^{-iHt}\ph_d\\ =&
\lim_{t\to\infty}e^{iHt} a^{\sharp} (h_{1,t} ) \dots a^{\sharp}
(h_{n,t})e^{-iHt}\ph.
\end{align*}
The second equation follows from
Theorem~\ref{theorem:many_photon_states_nonrel} and the last one
from the fact that, by
Theorem~\ref{theorem:many_photon_states_nonrel}, the convergence
as $d\to\infty$ is uniform in $t$.
\end{proof}
\end{theorem}

\section{Pseudo--Relativistic QED}\label{sec:halbrel} 
\subsection{Definition and Properties of the Model}
\label{subsec:def_mod_rel}

This section is devoted to scattering of photons at a single
electron in a pseudo-relativistic model of UV-cutoff QED. The
merits of this model have been discussed in
Section~\ref{sec:intro}. In units where the mass of the particle,
the speed of light, and Planck's constant are equal to one, the
Hamiltonian is given by
 \begin{equation}\label{eq:hamilt_rel}
H = \sqrt{(p+A(x))^{2} +1} + V(x)\otimes 1 + 1\otimes H_f ,
\end{equation}
and acts on the Hilbert space $\H=L^2(\R^3)\otimes \F$. Here
$p=-i\nabla_x$ and the quantized vector potential $A(x)$, the
field energy $H_f$, and the Fock space $\F$ are as in
Section~\ref{sec:nonrel}. To define $\sqrt{(p+A(x))^{2} +1}$ take
the Friedrichs' extension of the symmetric and positive operator
$(p+A(x))^{2} +1$ and then use the spectral theorem. The scalar
potential $V$ is defined by multiplication with the real-valued,
locally square integrable function $V(x)$ on $\R^3$. We assume $V$
is operator-bounded w.r.~to $\Omega(p)=\sqrt{p^2+1}$ with bound
less than 1. That is, there exist  constants $a<1$ and $b\in\R$
such that
\begin{equation}\label{eq:ass_on_V}
\| V\psi \| \leq a \| \Omega (p) \psi \| + b \| \psi
\|\makebox[8em]{for all}\psi\in C_0^{\infty}(\R^3).
\end{equation}
Using first the diagmanetic inequality and then the Schr\"odinger
representation of Fock space (see \cite{AHS} and \cite{FFrG}) one
shows that $V\otimes 1$ is also operator-bounded w.r.~to
$\Omega(p+A)=\sqrt{(p+A(x))^2+1}$ with bound less than 1. Hence
the operator given by Eq.~(\ref{eq:hamilt_rel}) is bounded from
below and may be self-adjointly realized by the Friedrichs'
extension. We denote this extension again by $H$. Moreover it
follows that $H_f$ and $\Omega(p+A)$ are form-bounded w.r.~to $H$.
For reasons given in Section~\ref{sec:nonrel} this is not enough
for the purpose of constructing scattering states with more than
one asymptotically free photon. We need to bound higher powers of
$H_f$ by powers of $H$.

\begin{lemma}\label{lemma:est_higher_ord_halbrel}
For any integer $m \geq 1$, the operators
\begin{itemize}
\item[i)] $[H_{f}^{m-1} , H ] (H+i)^{-m}$
\item[ii)] $H_{f}^{m} (H+i)^{-m}$
\end{itemize}
are bounded.
\end{lemma}

The proof of this lemma is long and difficult because of the
complicated form of the interaction between photons and electron.
It is sketched in Appendix \ref{sec:est_high_ord_rel}. By
Lemma~\ref{lemma:est_higher_ord_halbrel}, in particular
$H_f(H+i)^{-1}$ is bounded, which leads to the following
corollary.

\begin{corollary}\label{corollary:Omega_H_bound}
Assuming (\ref{eq:ass_on_V}) the operator $\Omega (p+A)
(H+i)^{-1}$ is bounded.
\end{corollary}

\begin{proof}
We use $\Omega$ as an abbreviation for the operator
$\Omega(p+A(x))$. For each $d > 0$ we have
\begin{equation*}
\begin{split}
\Omega (H+i)^{-1} = \; &H (H+i)^{-1} - H_f ( H+i)^{-1} \\ &- V(x)
(\Omega + d)^{-1} (\Omega +d ) (H+i)^{-1} ,
\end{split}
\end{equation*}
which shows that
\begin{equation*}
(1 + V(x) (\Omega + d)^{-1} ) \Omega (H+i)^{-1}\ \  \text{is
bounded} .
\end{equation*}
The corollary now follows, because, by (\ref{eq:ass_on_V}),
$\|V(x) (\Omega + d)^{-1}\| < 1$ for $d > 0$ sufficiently large.
\end{proof}

\subsection{A Sharp Propagation Estimate}
\label{subsec:prop_est_rel}

We begin to prove a propagation estimate similar to Theorem
\ref{theorem:prop_est_nonrel}. Unlike in the non--relativistic
case now the group velocity of the particle is always less than
the speed of light for finite energies. As a consequence
$a_{+}^{\sharp}(h)$ will exist on a dense subspace.

\begin{theorem}\label{theorem:prop_est_halbrel}
Let $f \in \mathcal{C}_{0}^{\infty} ( \R )$. Then, for each $\eps
> 0$ sufficiently small and for each $\mu > 1/2$, there exists a
constant $C$ such that, for all $\ph \in \mathcal{H}$,
\begin{equation*}
\int_{1}^{\infty} dt \, \frac{1}{t^{\mu}} \, \| \chi ( |x| \geq (1
- \eps)t ) e^{-iHt} f (H) \ph \|^{2} \leq C \, \| (1 + |x|)^{1/2}
f(H) \ph \|^2 .
\end{equation*}
\end{theorem}
{\em Remark.} The theorem actually holds for $\mu=0$ but for $\mu
>1/2$ the proof is easier and this result is sufficient for our
purposes.

\begin{proof}
This proof follows a similar pattern as the proof of
Theorem~\ref{theorem:prop_est_nonrel}. Some of the explanations
given in that proof are not repeated here. We may assume $f$ is
real-valued. Pick $h \in \mathcal{C}_{0}^{\infty} (\R)$ as in the
proof of Theorem~\ref{theorem:prop_est_nonrel} with the choice
$v=1-\eps$. That is, $h$ is non--decreasing, $0 \leq h \leq 1$,
$h(s)=0$ if $s \leq (1-2\eps)$ and $h(s)=1$ for $ s \geq
(1-\eps)$. Here $\eps
> 0$ and small. Again $\tilde{h}$ is defined by \(\tilde{h}(s)=\int_0^{s}d\tau
h^2(s)\) and obeys
\begin{equation}\label{eq:htildeh_rel}
\tilde{h}(s) \leq (s-(1-2\eps))h^2(s).
\end{equation}
The propagation observable is given by
\begin{equation*}
\phi (t) = - f (H)\,  t^{1-\mu} \tilde{h} (\expect{x} /t )\, f(H),
\end{equation*}
and the theorem follows if we show that
\begin{equation}\label{eq:reltodo}
\begin{split}
D\phi(t) &= -f(H)\{[iH,t^{1-\mu} \tilde{h}]+ ( \partial / \partial
t)(t^{1-\mu} \tilde{h})\}f(H)\\ & \geq \delta \, t^{-\mu} \,
f(H)h^2f(H) + (\text{integrable w.r.t. $t$})f(H)^2 ,
\end{split}
\end{equation}
for $t \in [T_0 , \infty )$, where $\delta>0$ and $T_0 > 1$ is
sufficiently large. By construction of $\tilde{h}$ and by
\eqref{eq:htildeh_rel}
\begin{equation}\label{eq:partial_tildeh_rel}
-\frac{\partial}{\partial t}(t^{1-\mu} \tilde{h}) \geq (1-2\eps )
\, t^{-\mu} \, h^2.
\end{equation}
To compute $[iH,\tilde{h}]$ note that $\tilde{h}(s)=s-c$ for
$s\geq 1-\eps$ where $c$ is constant. Hence there exists a
function $g\in C_0^{\infty}(\R)$ such that $\tilde{h}(s)=g(s)+s-c$
for $s>0$. This makes Lemma~\ref{lemma:pdc} applicable and leads
to
\begin{equation*}
\begin{split}
[iH , t^{1-\mu} \tilde{h} ] &= t^{1-\mu} \,  [ i \Omega ,
\tilde{h} ] \\ &=\frac{t^{1-\mu}}{2} \left( \nabla \Omega \cdot
\nabla \tilde{h} + \nabla \tilde{h} \cdot \nabla \Omega + O
(t^{-2}) \right).
\end{split}
\end{equation*}
Using next that $\nabla \tilde{h} = x/ \expect{x}\, t^{-1} h^2 $
and $[\nabla \Omega , h] = O(t^{-1})$ one finds
\begin{equation*}
[iH , t^{1-\mu} \tilde{h} ] =\frac{t^{-\mu}}{2} \, h \, \left(
\nabla \Omega \cdot \frac{x}{\expect{x}} + \frac{x}{\expect{x}}
\cdot \nabla \Omega \right) \, h + O(t^{-1-\mu})
\end{equation*}
by commuting one factor of $h$ to the left of $\nabla\Omega$,
respectively to the right of $\nabla\Omega$. Since $|x /
\expect{x}| \leq 1$ and by the Schwarz inequality we conclude
\begin{equation}\label{eq:scal_prod}
|\sprod{\ph_t}{f(H)[iH,t^{1-\mu}\tilde{h}]f(H)\ph_t}| \leq
t^{-\mu} \, \| h f(H)\ph_t\| \, \| \, |\nabla \Omega | \, h f(H)
\ph_t \| + O(t^{-1-\mu})f(H)^2.
\end{equation}
To estimate the factor $\| \, |\nabla \Omega | \, h f(H) \ph_t \|$
we pick $g\in C_0^{\infty}(\R)$ with $gf=f$ and then commute $h$
with $g(H)$ using $[g(H) , h ] =O(t^{-1})$. This shows that
\begin{equation}\label{eq:f_H_comm}
f(H)\, h \, | \nabla \Omega |^{2} \, h \, f(H) \leq f(H)\,h\,
g(H)\, | \nabla \Omega |^{2} \, g(H)\,h\, f(H) + O(t^{-1})f(H)^2.
\end{equation}
Next write $| \nabla \Omega |^2 = 1 - \Omega^{-2}$ and let $P=
\chi(H\in\supp\, g)$. From $1 \leq \Omega \leq \alpha H + \beta $
for some constants $\alpha , \beta $, it follows that $\Omega^{-1}
\geq (\alpha H + \beta )^{-1}$, and hence that
\begin{equation*}
P \frac{1}{\Omega^2} P \geq (P\frac{1}{\Omega}P)^2 \geq ( 8 \eps
-16 \eps^2 ) P ,
\end{equation*}
if $\eps > 0$ is small enough. These remarks in conjunction with
(\ref{eq:f_H_comm}) show that
\begin{equation}\label{eq:1_4eps}
f(H) \, h \, |\nabla \Omega |^2 \, h \, f(H) \leq (1 -4\eps )^2
 \, f(H) \, h^{2} \, f(H) + O(t^{-1})f(H)^2,
\end{equation}
where we commuted $h$ and $g(H)$ once again. We now insert this
result in (\ref{eq:scal_prod}) and we get
\begin{equation*}
\begin{split}
|\sprod{\ph_t}{f(H)[iH,t^{1-\mu}\tilde{h}]f(H)\ph_t}| \leq \;
&t^{-\mu} (1- 4\eps ) \, \| h f(H) \ph_t \|^2 \\ &+ O
(t^{-1/2-\mu}) \, \| f(H)\ph \|^2 .
\end{split}
\end{equation*}
From the last equation, and from (\ref{eq:partial_tildeh_rel}), it
follows that
\begin{equation*}
\begin{split}
\sprod{\ph_t}{D\phi (t) \ph_t} \geq \; &\{ (1 - 2 \eps ) - (1- 4
\eps) \} \, t^{-\mu} \, \| h f(H) \ph_t \|^2 \\ &+ C \, t^{-\mu -
1/2} \, \| f(H)\ph \|^{2}.
\end{split}
\end{equation*}
This implies (\ref{eq:reltodo}), with $\delta=2\eps>0$, and
completes the proof of the theorem.
\end{proof}

\subsection{Existence of Asymptotic Field Operators}
\label{subsec:existence_rel}

Following the general strategy outlined in the introduction we
next use the propagation estimate to prove existence of $a_{+}^*
(h) $ by the Cook method. Existence of scattering states with more
than one photon then follows with the help of Lemma
\ref{lemma:est_higher_ord_halbrel}. The proofs are similar as in
the non--relativistic case and therefore kept short.

The following lemma will serve us to compute commutators with
$\Omega =\sqrt{(p+A)^{2} + 1}$.
\begin{lemma} \label{lemma:commutator}
Let $B$ be an operator on $\mathcal{H}$. Then
\begin{equation}\label{eq:commut_omega}
[ \Omega , B ] = \frac{1}{\pi} \int_{1}^{\infty} dy \,
\frac{\sqrt{y-1}}{y+(p+A)^2} \, [(p+A)^2,B]\,\frac{1}{y+(p+A)^2}.
\end{equation}
\end{lemma}
\begin{proof}
Using the representation
\begin{equation*}
\Omega^{-1} =( 1 + (p+A)^2)^{-1/2} =\frac{1}{\pi}
\int_{1}^{\infty} dy \, \frac{1}{\sqrt{y-1}} \, \frac{1}{y +
(p+A)^2},
\end{equation*}
we get
\begin{equation*}
\begin{split}
[ \Omega , B ] &= \Omega^2 \, [\Omega^{-1} , B ] +[ (p+A)^2 , B ]
\Omega^{-1} \\ &=\frac{1}{\pi} \int_{1}^{\infty}  dy \,
\frac{1}{\sqrt{y-1}} \, \left\{ - ((p+A)^{2} +1) (y +
(p+A)^2)^{-1} + 1 \right\} \\ &\qquad \qquad \times [(p+A)^2 , B ]
(y + (p+A)^2)^{-1} .
\end{split}
\end{equation*}
The statement of the lemma follows from the last equation since
the factor in the braces, on the r.h.s. of the last equation, is
equal to $ (y -1 ) \, (y + (p+A)^2 )^{-1}$.
\end{proof}

The following proposition establishes existence of $a_{+}^* (h)$
on the subspace $ \cup_{d > 0} \chi ( H \leq d ) \mathcal{H}$ and
for $h \in \mathcal{C}_{0}^{\infty} (\R^3 \backslash \{ 0\}; \C^2
)$.
\begin{proposition}\label{proposition:existenz}
Suppose $f \in \mathcal{C}_{0}^{\infty}( \R )$ and $h \in
\mathcal{C}_{0}^{\infty} (\R^{3} \backslash \{ 0 \})$. Then, for
all $\ph \in \mathcal{H}$, the limit
\begin{equation}\label{eq:to_prove_rel}
\lim_{t \to \infty} e^{iHt} a^{\sharp} (h_{t}) e^{-iHt} f(H) \ph
\end{equation}
exists.
\end{proposition}
\begin{proof}
Since $e^{iHt} a^{\sharp} (h_t) e^{-iHt} f(H) $ is bounded
uniformly in $t$ it suffices to prove existence of
(\ref{eq:to_prove_rel}) for $\ph \in D(\expect{x}^{1/2})$. We only
consider creation operators, the proof for annihilation operators
is similar. For given $\ph \in \mathcal{H}$ let
\begin{equation*}
\ph (t) = e^{iHt}\, a^{*} (h_{t}) \, e^{-iHt} \, f \, \ph ,
\end{equation*}
where $f = f(H)$. By Cook's argument the existence of the limit
(\ref{eq:to_prove_rel}) follows if
\begin{equation*}
\int_{1}^{\infty} dt \, \| \frac{d}{dt} \ph (t) \| < \infty .
\end{equation*}
In the following we will use the notation $\ph_t = e^{-iHt} \ph$.
A straightforward computation shows that
\begin{equation}\label{eq:d_dt_ph_t_rel}
\begin{split}
\frac{d}{dt} \ph (t) &= e^{iHt} \, [i\Omega \, , a^{*} (h_{t})]\,
f \ph_{t} \\ &=\frac{1}{\pi} \int_{1}^{\infty} dy \, \sqrt{y-1} \,
e^{iHt} R(y) [ i(p+A)^{2},a^{*}(h_{t})] R(y) f \ph_{t},
\end{split}
\end{equation}
where we expanded the commutator with $\Omega$ using Lemma
\ref{lemma:commutator}, and where $R(y) = (y + (p+A)^{2})^{-1}$.
Since
\begin{equation*}
[ (p+A)^{2} , a^{*} (h_{t})] =  \sqrt{2} (G_{x} , h_{t}) \cdot (p
+ A ) ,
\end{equation*}
we see that
\begin{equation}\label{eq:est_sprod}
\| \frac{d}{dt}  \ph (t)\| \leq \frac{\sqrt{2}}{\pi}
\int_{1}^{\infty} dy \, \sqrt{y-1} \, \| R(y) \| \, \| (G_{x} ,
h_{t}) \cdot (p+ A)\, R(y) f \ph_{t} \|.
\end{equation}
Next choose $\chi_{1} \in \mathcal{C}_{0}^{\infty} (\R)$, with $0
\leq \chi_{1} \leq 1$, $\chi_{1} (s) =0$ if $ s \leq 1- 2 \eps $
and $\chi_{1} (s) =1$ for $s \geq 1-\eps$. Here $\eps > 0$ and so
small that the propagation estimate (Theorem
\ref{theorem:prop_est_halbrel}) holds for the cut--off function
$(s+i) f(s)$ and $\eps$. Let $\chi_{2}^2 = 1 -\chi_{1}^2$ and let
$\chi_{1}=\chi_{1} ( |x| / t )$, $\chi_{2} = \chi_{2} ( |x| / t )$
henceforth. Inserting
\begin{equation*}
(G_{x} , h_{t} ) = \chi_{1}^{2} \, (G_{x} , h_{t}) + \chi_{2}^{2}
\, (G_{x} , h_{t})
\end{equation*}
in the r.h.s. of (\ref{eq:est_sprod}), we find
\begin{equation}\label{eq:est_sprod2}
\| \frac{d}{dt} \ph (t) \| \leq \frac{\sqrt{2}}{\pi} \sum_{k=1,2}
\int_{1}^{\infty} dy \, \frac{\sqrt{y-1}}{y} \, \| (G_{x} , h_{t})
\chi_k \| \, \| \chi_k ( p+A) R(y) f \ph_{t} \| .
\end{equation}
Consider first the term with $k=2$. Since
\begin{equation*}
\| (p+ A) R(y) f \| \leq  \| (p +A) \Omega^{-1} \| \, \| R(y)
\Omega  f \| \leq C / y ,
\end{equation*}
and since, by a stationary phase argument,
\begin{equation*}
\| \chi_{2} \, (G_{x},h_{t}) \| \leq \frac{C_n}{t^n},
\end{equation*}
for any $n \geq 1$, the term with $k=2$ on the r.h.s. of
(\ref{eq:est_sprod2}) is bounded by
\begin{equation*}
\const  \int_{1}^{\infty} dy \, \frac{\sqrt{y-1}}{y^{2}} \, t^{-2}
\leq \frac{\const }{t^2},
\end{equation*}
and is therefore integrable w.r.t. $t$. Consider now the term with
$k=1$. We first note that
\begin{equation}\label{eq:comm_chi_pR}
\begin{split}
\| \chi_1 (p+A) R(y) f \ph_{t} \| &\leq \| (p+A) R(y) (H+i)^{-1}
\| \, \| \chi_1 \tilde{f} \ph_{t} \| + \frac{C}{y \, t} \\ &\leq
\frac{C}{y} \, \| \chi_1 \tilde{f} \ph_{t} \| + \frac{C}{y \, t},
\end{split}
\end{equation}
where $\tilde{f} = (H+i)f$. To see this write $f=(H+i)^{-1}
\tilde{f}$ and then commute the operator $(p+A) R(y) (H+i)^{-1}$
to the left of $\chi_1$. The second term on the r.h.s. of
(\ref{eq:comm_chi_pR}) arises from the commutator of these two
operators. Since, by Lemma \ref{lemma:phot_disp}, $\| \chi_1
(G_{x_j} , h_t )\|\leq C / t$, it follows that the term with $k=1$
on the r.h.s. of (\ref{eq:est_sprod2}) is bounded by $\const \,
t^{-1}  \, \| \chi_1 \tilde{f} \ph_t \| + \const \, t^{-2}$. The
term proportional to $t^{-2}$ is clearly integrable w.r.t. $t$ and
by the Schwarz inequality
\begin{equation}
\int_{1}^{\infty} dt \, \frac{1}{t} \, \| \chi_1 \tilde{f} \ph_t
\| \leq \left( \int_1^{\infty} dt \, \frac{1}{t^{1+2\delta}}
\right)^{\frac{1}{2}} \left( \int_1^{\infty} dt \,
\frac{1}{t^{1-2\delta}} \, \| \chi_1 \tilde{f} \ph_t \|^2
\right)^{\frac{1}{2}}.
\end{equation}
If we choose $\delta \in (0 , 1/4)$ this is finite by Theorem
\ref{theorem:prop_est_halbrel} with $\mu = 1 - 2 \delta > 1/2$,
because $\tilde{f} \ph \in D(\expect{x}^{1/2})$ by Lemma
\ref{lemma:invar}.
\end{proof}

Using Proposition \ref{proposition:existenz} and an approximation
argument we now extend existence of $a_{+}^* (h) \ph $ to $h \in
L_{\omega}^{2} (\R^3 ; \C^2 )$ and to $\ph \in D((H+i)^{1/2})$.

\begin{theorem}\label{theorem:defandprop}
Suppose $g,h \in L_{\omega}^{2}(\R^3 ; \C^2 )$. Then
\begin{enumerate}
\item[i)]
\begin{equation*}
a_{+}^{\sharp} (h) \ph = \lim_{t \to \infty} e^{iHt} a^{\sharp}
(h_{t}) e^{-iHt} \ph
\end{equation*}
exists for all $\ph \in D((H+i)^{1/2})$.
\item[ii)]
\begin{equation*}
[ a_{+} (g) , a_{+}^{*} (h) ] = (g,h) \qquad \text{and} \qquad
[a_{+}^{\sharp}(h) , a_{+}^{\sharp}(g)] = 0 ,
\end{equation*}
in form-sense on $D((H+i)^{1/2})$.
\item[iii)]
If $\omega h \in L^2 (\R^3 ; \C^2 )$, in addition to $h \in
L^{2}_{\omega} (\R^3
; \C^2 )$, then
\begin{equation*}
[H , a_{+}^{*} (h) ] = a_{+}^{*} (\omega h), \qquad  [ H, a_{+}
(h) ] = -a_{+} (\omega h),
\end{equation*}
in form-sense on $D(H)$, and
\begin{equation*}
\phi_{+} (h) = \frac{1}{\sqrt{2}} (a_{+}^{*} (h) + a_{+} (h))
\end{equation*}
is essentially self adjoint on $ \cup_{d > 0} \chi (H \leq d)
\mathcal{H}$.
\item[iv)] Set $M= \sup \{ |k| : h(k) \neq 0 \}$ and
$m=\inf \{ |k| : h(k) \neq 0\}$. Then
\begin{equation*}
\begin{split}
a_{+}^{*} (h) \Ran \chi (H \leq E) &\subset \Ran \chi (H\leq E+M)
\\ a_{+} (h) \Ran \chi (H \leq E) &\subset \Ran \chi (H\leq E-m )
.
\end{split}
\end{equation*}
\end{enumerate}
\end{theorem}

\begin{proof}
\begin{enumerate}
\item[i)] Follows from $\mathcal{C}_{0}^{\infty} (\R^{3} \backslash \{
0 \} ) \subset L_{\omega}^{2} (\R^{3})$ dense, from $\| a^{\sharp}
(h) (H+i)^{-1/2} \| \leq C \|h \|_{\omega}$ and because $\cup \chi
(H \leq d ) \mathcal{H}$ is a form-core for $H$.
\item[ii)] Follows from i) and the CCR for $a(h)$ and $a^{*}
(h)$.
\item[iii)] For $\ph , \psi \in D(H)$, by ii),
\begin{equation*}
\begin{split}
\sprod{\ph}{ [H, a_{+}^{*} (h)] \psi} &= \lim_{t \to \infty}
\sprod{\ph_{t}}{ [H , a^{*} (h_{t})]\psi_{t}} \\ &= \lim_{t \to
\infty} \sprod{\ph_{t}}{ [\Omega , a^{*} (h_{t})] \ph_{t}} +
\sprod{\ph }{a_{+}^{*}(\omega h )\psi }.
\end{split}
\end{equation*}
Now, by Lemma \ref{lemma:commutator},
\begin{equation*}
[\Omega , a^{*} (h_{t})] = \frac{\sqrt{2}}{\pi} \int_1^{\infty} dy
\, \frac{\sqrt{y-1}}{y+(p+A)^2} (G_{x},h_{t}) \cdot (p+A)
\frac{1}{y+(p+A)^2}.
\end{equation*}
Since $\sup_{x} | (G_{x} , h_{t} )| \to 0$ for $t \to \infty$, by
Lemma \ref{lemma:phot_disp}, it follows that $ [ \Omega , a^{*}
(h_{t}) ] \, \Omega^{ -1} \to 0$ as $t \to \infty$. This proves
the first pull through formula, the proof of the second is
similar. The essential self-adjointness of $\phi_{+} (h)$ follows
from $[H,\phi_{+} (h) ] = - i\phi_{+} (i\omega h)$ and from
Nelson's commutator theorem (\cite{RSII}, Theorem X.37), since
$\phi_{+} (h)$ is symmetric and bounded w.r.t. $(H+i)^{1/2}$.
\item[iv)]Follows in the same way as in the non--relativistic case (see Theorem
\ref{theorem:scatt_states_nonrel}, iv)).
\end{enumerate}
\end{proof}

The following theorem, on the existence of scattering states with
an arbitrary number of asymptotically free photons, together with
Theorem \ref{theorem:defandprop} is the main result of this
section.

\begin{theorem}\label{theorem:many_phot_states_halbrel}
Suppose $h_i \in L_{\omega}^{2} (\mathbb{R}^{3})$, for all $i \in
\{1,\dots ,n \}$ and $\ph \in D((H+i)^{n/2})$. Then $\ph \in
D(a_{+}^{\sharp} (h_{1}) \cdots a_{+}^{\sharp} (h_{n}))$,
\begin{equation}\label{eq:limit_many_rel}
a_{+}^{\sharp} (h_{1}) \dots a_{+}^{\sharp} (h_{n}) \ph = \lim_{t
\to \infty} e^{iHt} a^{\sharp} (h_{1,t}) \dots a^{\sharp}
(h_{n,t}) e^{-iHt}\ph ,
\end{equation}
and
\begin{equation}\label{eq:bound_many_phot_halbrel}
\| a_{+}^{\sharp} (h_{1}) \dots a_{+}^{\sharp} (h_{n})
(H+i)^{-n/2} \| \leq C \|h_{1} \|_{\omega} \cdots \|h_{n}
\|_{\omega} .
\end{equation}
\end{theorem}

The proof follows the same lines as in the non--relativistic case
and is given in the Appendix \ref{sec:est_high_ord_rel}.
Technically the most difficult part is to show Lemma
\ref{lemma:est_higher_ord_halbrel} which, together with
Theorem~\ref{theorem:defandprop}, is the main ingredient.

\section{Weyl Operators} 
\label{sec:Weyl}

The purpose of this section is to establish existence of the
asymptotic Weyl operators \(W_{+}(h)=
s-\lim_{t\to\infty}e^{iHt}W(h_t)e^{-iHt}\) and to show that
$W_{+}(h)= e^{i\phi_{+}(h)}$. We will treat the relativistic and
the non-relativistic model simultaneously, as most of the
arguments are independent of the model. However, in the
non-relativistic case we assume that $\Sigma=\infty$. The Weyl
operators are defined by
\begin{equation*}
W(h) = e^{i\phi(h)},\hspace{3em} h\in L^2(\R^3;\C^2)
\end{equation*}
where $\phi(h)$ denotes the closure of
\(1/\sqrt{2}(a(h)+a^{*}(h))\), which is self-adjoint. They obey
the Weyl relations
\begin{equation}\label{Weyl}
W(g+h) = W(g) W(h) e^{i/2\Ima(g,h)}.
\end{equation}

\begin{theorem}\label{theorem:weyl}
Suppose $H=H^{rel}$ or $H=H^{nr}$ on the appropriate Hilbert space
and assume $\Sigma=\infty$ in the second case. If $g,h\in
L^2_{\omega}$ then
\begin{itemize}
\item[(i)] The strong limits \(W_{+}(h)=s-\lim_{t\to\infty}e^{iHt}W(h_t)e^{-iHt}\) exist.
\item[(ii)] The asymptotic Weyl operators obey the
  Weyl relations \[ W_{+}(g+h) = W_{+}(g) W_{+}(h) e^{i/2\Ima(g,h)}.\]
\item[(iii)] The mapping $\R\ni s\mapsto W_{+}(sh)$ defines a strongly
  continuous, one parametric group of unitary operators generated by the closure of $\phi_{+}(h)$.
\end{itemize}
\end{theorem}

\begin{proof}
(i) Suppose first $h\in C_0^{\infty}(\R^3\backslash\{0\})$ and
pick $f\in C_0^{\infty}(\R)$. Then existence of
\begin{equation}\label{eq:asy_Weyl}
\lim_{t\to\infty}e^{iHt}W(h_t)e^{-iHt}f(H)\ph
\end{equation}
is proved as in \eqref{proposition:exist_nonrel} and
\eqref{proposition:existenz} with only small modifications: rather
than $[(p+A(x))^2,a^{*}(h_t)]=\sqrt{2} \, (p + A(x))\cdot ( G_{x}
, h_t )$ as in equations \eqref{eq:d_dt_ph_t} and
\eqref{eq:d_dt_ph_t_rel} one now has $[(p+A(x))^2,W(h_t)]$ for
which we use the formula
\begin{equation}
[(p+A(x))^2,W(h_t)] =
-W(h_t)\left\{2\Ima(G_x,h_t)\cdot(p+A(x))-[\Ima(G_x,h_t)]^2\right\}.
\end{equation}
New is the factor $W(h_t)$ in front, which does not affect the
subsequent estimates since $\|W(h_t)\| = 1$, and the additional
term $[\Ima(G_x,h_t)]^2$ which is dealt with similarly as the
first term in braces.

Next if $\ph=\chi(D\leq d)\ph$ and $h\in L^2_{\omega}$ pick $f\in
C_0^{\infty}(\R)$ with $f(H)\ph=\ph$. Existence of
\eqref{eq:asy_Weyl} then follows by an approximation argument
using a sequence $\{h_n\}\subset
C_0^{\infty}(\R^3\backslash\{0\})$ with $\|h_n-h\|_{\omega}\to 0$
and Lemma~\ref{Weyl_tools}. This establishes existence of
$W_{+}(h)$ as a strong limit on the dense subspace
$\bigcup_{d>0}\chi(H\leq d)\H$. Statement (i) now follows by an
other approximation argument.

Statement (ii) follows easily from the Weyl relations \eqref{Weyl}
and from (i).

(iii). From Lemma~\ref{Weyl_tools} and (i) it follows that the
mapping \(L^2_{\omega}\ni h\mapsto W_{+}(h)(H+i)^{-1/2}\) is
continuous. This implies that $L^2_{\omega}\ni h\mapsto W_{+}(h)$
is strongly continuous and hence so is \(\R\ni s\to W_{+}(sh)\).
Unitarity and the group structure follow from (i) and (ii).

Finally, because of the essential self-adjointness of
$\phi_{+}(h)$ on $D(H)$, it only remains to show that
\begin{equation*}
\lim_{s\to 0}\frac{1}{s}(W_{+}(sh)-1)\ph = \phi_{+}(h)\ph
\end{equation*}
for all $\ph\in D(H)$. By (i),
Theorem~\ref{theorem:scatt_states_nonrel} in the non-relativistic
case, and by Theorem~\ref{theorem:defandprop} in the relativistic
case, this is equivalent to
\begin{equation*}
\lim_{s\to
\infty}\lim_{t\to\infty}e^{iHt}\frac{1}{s}[W(sh_t)-1-i\phi(sh_t)]e^{-iHt}\ph=0.
\end{equation*}
But this follows from
\begin{equation*}
\|[W(sh_t)-1-i\phi(sh_t)](H+i)^{-1}\| \leq
\frac{1}{2}\|\phi(sh)^2(H+i)^{-1}\|\leq Cs^2
\end{equation*}
which holds uniformly in $t$. Here we used Lemma~\ref{Weyl_tools}
again and Lemma~\ref{lemma:ann_cre_bound}.
\end{proof}

\begin{lemma}\label{Weyl_tools}
\begin{align*}
(i) \quad & \|[W(h)-1-i\phi(h)]u\| \leq \frac{1}{2}\|\phi(h)^2
u\|\\ (ii) \quad & \|[W(h_1)-W(h_2)]u\| \leq \|h_1-h_2\|
\|h_1+h_2\| \|u\| + 2\|h_1-h_2\|_{\omega}\|(H_f+1)^{1/2}u\|
\end{align*}
\end{lemma}

\begin{proof}
(i) follows from $|e^{it}-1-it|\leq t^2/2$ and the spectral
theorem. (ii) follows from
\begin{equation}
W(h_1)-W(h_2) = W(h_1)\left(1-e^{i/2\Ima(h_1,h_2)}\right) + W(h_1)
e^{i/2\Ima(h_1,h_2)}(1-W(h_2-h_1)).
\end{equation}
\end{proof}

\appendix 

\section{Estimating Field Operators}\label{sec:a_bound}

Recall from Section~\ref{sec:nonrel} that $L_{\omega}^{2} (\R^3 ;
\C^2 ) = L^2 (\R^3 , (1+ 1/ |k|) \, dk)\otimes \C^2$ and that $\|
\, . \, \|_{\omega}$ denotes the norm of this space.

\begin{lemma}\label{lemma:ann_cre_bound}
Suppose $h_i \in L_{\omega}^2 (\R^3 ; \C^2 )$ for $i=1,\dots , n$.
Then
\begin{equation*}
\| a^{\sharp} (h_1) \dots a^{\sharp} (h_n) (H_f +1)^{-n/2} \| \leq
C_n \, \prod_{i=1}^{n} \,  \|  h_i \|_{\omega},
\end{equation*}
with a finite constant $C_n$.
\end{lemma}
\begin{proof}
Let $\bar{\omega} (k) = \min \{ |k| , 1\}$, $d\Gamma (\omega) =
H_f$ and
\begin{equation*}
d\Gamma (\bar{\omega}) = \sum_{\lambda=1,2} \int dk \,
\bar{\omega} (k) a_{\lambda}^* (k) a_{\lambda} (k).
\end{equation*}
Since $\bar{\omega} \leq \omega$ we have $\| (d\Gamma
(\bar{\omega}) +1)^{n/2} (d\Gamma(\omega) +1)^{-n/2}\| \leq 1$ and
hence it suffices to show that the lemma holds with $H_f$ replaced
by $d\Gamma (\bar{\omega})$. For the case $n=1$ see e.g.
\cite{BFS1}. If $n>1$ let $\Lambda = (d\Gamma (\bar{\omega}) +1)$
and note that
\begin{equation}\label{eq:a_dots_a}
a^{\sharp} (h_1) \dots a^{\sharp} (h_n) \Lambda^{-n/2} =
\prod_{k=1}^{n} \Lambda^{(k-1)/2} a^{\sharp} (h_k) \Lambda^{-k/2}.
\end{equation}
Next, in each factor $\Lambda^{(k-1)/2} a^{\sharp} (h_k)
\Lambda^{-k/2}$ commute as many factors of $\Lambda$ to the right
as possible. For $k$ even one gets
\begin{equation}\label{eq:k_even}
\Lambda^{(k-1)/2} a^{\sharp} (h_k) \Lambda^{-k/2}= \Lambda^{-1/2}
a^{\sharp} (h_k) +  \Lambda^{-1/2} [ \Lambda^{k/2} , a^{\sharp}
(h_k)] \Lambda^{-k/2}
\end{equation}
and a similar formula holds for odd $k$. From (\ref{eq:k_even}),
the commutator expansion (\ref{eq:comm_exp1}) for $[\Lambda^{k/2}
, a^{\sharp} (h) ]$, and $\ad_{\Lambda}^{l} (a^{\sharp} (h) ) =
(\pm1)^{l} a^{\sharp} (\bar{\omega}^l h)$ it follows that
\begin{equation*}
\| \Lambda^{(k-1)/2} a^{\sharp} (h_k) \Lambda^{-k/2} \| \leq C_k
\| h_k \|_{\omega}
\end{equation*}
which, together with (\ref{eq:a_dots_a}) proves the lemma.
\end{proof}

In particular $A(x)^2$ is form-bounded with respect to $H_f+1$. As
a consequence we have the following lemma.

\begin{lemma}\label{lemma:eps-bound}
Suppose $H = H^{nr}$, and that for each $\eps>0$ there exists a
$C_{\eps}>0$ such that $V_{-}\leq \eps(-\Delta)+C_{\eps}$ in
form-sense. Then
\begin{equation}
V_{-} \leq \eps H +D_{\eps}\hspace{3em}\makebox[5em]{for
all}\eps>0
\end{equation}
with another constant $D_{\eps}$.
\end{lemma}

\begin{proof}
It is enough to show that $-\Delta=\sum_{i=i}^{N}p_j^2$ is
form-bounded with respect to $H$. Since \( A(x)^{2} \leq
C(H_f+1)\) by lemma~\ref{lemma:ann_cre_bound}, and from $p^2\leq
2(p+A(x))^2+2A(x)^2$ it follows that
\begin{equation}
\sum_{j=1}^N p_j^2\leq a(H+V_{-}) + b
\end{equation}
for some $a,b> 0$. Combined with the assumption on $V_{-}$ this
shows that $\sum_{i=i}^{N}p_j^2$ is form-bounded with respect to
$H$ and hence proves the lemma.
\end{proof}

\section{Commutator Estimates}
In this section we collect some techniques which are useful to
compute and to estimate commutators and which are frequently used
in this paper.

\subsection{Commutator Expansions}
\label{sec:exp_comm}\label{sec:commut_estim}

First we recall that, given two operators $A,B$ acting on the same
space,
\begin{align}
[A^{n} , B] &= \sum_{l=1}^{n} \, \left( \begin{array}{c} n \\ l
\end{array}\right) \ad_{A}^{l} (B) A^{n-l} \label{eq:comm_exp1} \\
\ad_{A}^{n} (BC) &= \sum_{l=0}^{n} \, \left( \begin{array}{c} n \\
l
\end{array}\right) \ad_{A}^{l} (B) \ad_{A}^{n-l} (C) \label{eq:comm_exp2}
\end{align}
where $\ad_{A}^{n} (B)$ is defined by $\ad_{A}^{0}(B) =B$ and
$\ad_{A}^{n} (B) = [ A , \ad_{A}^{n-1} (B)]$ and where $C$ is a
third operator.

In the case where the two operators to be commuted are defined by
multiplication with functions in position and momentum space
respectively, the following lemma is very useful.

\begin{lemma}\label{lemma:pdc}
Suppose $f\in \mathcal{S}(\R^d)$, $g\in \mathcal{C}^{2}(\R^{d})$
and $\sup_{|\alpha|=2}\|\partial^{\alpha}g\|_{\infty}<\infty$. Let
$p=-i\nabla$. Then
\begin{align*}
i[g(p),f(x)]&= \nabla f (x)\cdot \nabla g (p) + R_{1}\\ &= \nabla
g (p) \cdot \nabla f (x) + R_{2}
\end{align*}
where, for $j=1,2$, \[\Vert R_{j}\Vert\leq C
\sup_{|\alpha|=2}\|\partial^{\alpha}g\|_{\infty} \int
dk\,|k|^2|\hat{f}(k)|.\]
\end{lemma}
\begin{proof}
Let \(f(x)=\int dk\, e^{ikx}\hat{f}(k)\). The first equation
follows from
\begin{equation}
\begin{split}\label{pdc1}
g(p)e^{ikx}-e^{ikx}g(p) &=
e^{ikx}\left[e^{-ikx}g(p)e^{ikx}-g(p)\right]\\ &=
e^{ikx}[g(p+k)-g(p)]
\end{split}
\end{equation}
and Taylor's formula
\begin{align*}
g(p+k)-g(p) = \nabla g (p) \cdot k \\
 &+ \int_0^1 dt\,
 (1-t) \sum_{|\alpha|=2}(\partial^{\alpha}g)(p+tk)k^{\alpha}.
\end{align*}
To obtain the second equation write
\[ g(p)e^{ikx}-e^{ikx}g(p) = -[g(p-k)-g(p)]e^{ikx}\]
instead of \eqref{pdc1}.
\end{proof}

Note that the Lemma basicly follows from $[p_i , x_j ] =
-i\delta_{ij}$. In particular the statement also holds for $p$
replaced by $p+A(x)$.

\subsection{Helffer--Sj\"ostrand Functional Calculus}
\label{sec:hsfuncal} Suppose $f\in C_0^{\infty}(\R;\C)$ and $A$ is
a self-adjoint operator. A convenient representation of $f(A)$ is
then given by
\begin{equation*}
f (A) = -\frac{1}{\pi} \int
dxdy\,\frac{\partial\tilde{f}}{\partial\bar{z}} (z)\, (z -A)^{-1},
\hspace{2em}z=x+iy,
\end{equation*}
which holds for any extension $\tilde{f}\in C_0^{\infty}(\R^2;\C)$
of $f$ with $|\partial_{\bar{z}}\tilde{f}|\leq C|y|$,
\begin{equation}\label{eq:alman}
\tilde{f} (z) = f(z) \makebox[5em]{and}
\frac{\partial\tilde{f}}{\partial\bar{z}} (z)
=\frac{1}{2}\left(\frac{\partial f}{\partial x}+ i\frac{\partial
f}{\partial y} \right)(z)= 0 \makebox[7em]{for all}z \in \R.
\end{equation}
Such a function $\tilde{f}$ is called an {\em almost analytic
extension} of $f$. A simple example is given by \(\tilde{f} (z) =
(f(x) + i y f^{\prime} (x)) \, \chi (z)\) where $\chi \in
C_0^{\infty} (\R^2)$ and $\chi = 1$ on some complex neighborhood
of $\supp f$. Sometimes we need faster decay of
$|\partial_{\bar{z}}\tilde{f}|$ as $|y|\to 0$ in the form
$|\partial_{\bar{z}}\tilde{f}|\leq C |y|^n$. In that case we work
with the almost analytic extension
\begin{equation*}
\tilde{f}(z) = \left(\sum_{k=0}^n
f^{(k)}(x)\frac{(iy)^k}{k!}\right)\chi(z)
\end{equation*}
where $\chi$ is as above. For more details and extensions of this
functional calculus the reader is referred to \cite{HunSig} or
\cite{Dav}.

\section{Invariance of Domains}

The following lemma provides us with a dense subspace on which
$(1+x^2)^{1/4}f(H)$ is defined. This is important in the proofs of
the Propositions \ref{proposition:exist_nonrel} and
\ref{proposition:existenz}.

\begin{lemma}\label{lemma:invar}
Let $H=H^{rel}$ or $H=H^{nr}$ and let $\expect{x}=(1+x^2)^{1/2}$,
where $x^2 = \sum_{j=1}^{N} x_j^{2}$ in the non-relativistic case.
If $f\in C_0^{\infty}(\R)$ then $[\expect{x}^{1/2} , f(H) ]$ is a
bounded operator and hence
\[ f(H) D(\expect{x}^{1/2}) \subset D(\expect{x}^{1/2}).  \]
\end{lemma}

\begin{proof}
We only prove the lemma for the case $H=H^{rel}$. The case
$H=H^{nr}$ is similar and easier.

The second statement follows from the first one by an argument
using the self-adjointness of $\expect{x}^{1/2}$. To show that
$[\expect{x}^{1/2} , f(H) ]$ is a bounded operator let $\tilde{f}$
be an almost analytic extension of $f$ with
$|\partial_{\bar{z}}\tilde{f}|\leq C |y|^2$ (see Appendix
\ref{sec:hsfuncal}). Then
\begin{equation}\label{eq:[x,f]_rel}
\begin{split}
[\expect{x}^{1/2} , f(H) ] &=-\frac{1}{\pi} \int dudv \,
\partial_{\bar{w}} \tilde{f} (w) \, [\expect{x}^{1/2} , (w
-H)^{-1} ] \\ &=-\frac{1}{\pi} \int dudv \, \partial_{\bar{w}}
\tilde{f}(w) \,  (w-H)^{-1} [\expect{x}^{1/2} , \Omega ] \,
(w-H)^{-1},
\end{split}
\end{equation}
where $w=u+iv$. By Lemma~\ref{lemma:commutator}
\begin{equation}\label{eq:[x,omega]}
\begin{split}
[\expect{x}^{1/2} , \Omega ] = \frac{i}{2 \pi} &\int_1^{\infty} dy
\, \sqrt{y-1} \, (y + (p+A)^2)^{-1} \\ &\times \left\{ (p+A) \cdot
\frac{x}{\expect{x}^{3/2}} + \frac{x}{\expect{x}^{3/2}} \cdot
(p+A) \right\} (y+(p+A)^2)^{-1}.
\end{split}
\end{equation}
Insert this into (\ref{eq:[x,f]_rel}). From
\begin{equation*}
\begin{split}
\| (w -H)^{-1} &(y +(p+A)^{2})^{-1} (p+A) \| \\ &\leq \| (w
-H)^{-1} \Omega \| \| (y+(p+A)^2)^{-1}\| \| \Omega^{-1} (p+A) \|
\\ &\leq \frac{C}{y} (\frac{1 + |w|}{|v|}),
\end{split}
\end{equation*}
and the boundedness of $x / \expect{x}^{3/2}$ it then follows that
\begin{equation*}
\| \, [ \expect{x}^{1/2} , f(H) ] \, \| \leq \const \int du dv \,
| \partial_{\bar{w}} \tilde{f} | \, (\frac{1+|w|}{|v|^{2}}) \,
\int_1^{\infty} dy \, \frac{\sqrt{y-1}}{y^2}.
\end{equation*}
This is finite by Appendix~\ref{sec:hsfuncal} and hence
$[\expect{x}^{1/2} , f(H)]$ is a bounded operator.
\end{proof}

\section{Non--Relativistic QED: Higher Order Estimates}
\label{sec:est_high_ord_nonrel} In this section we want to prove
Lemma \ref{lemma:estim_higher_ord_nonrel}, which ensures the
boundedness of $H_f^n$ w.r.t. $H^n$, where $H_f$ and $H$ are the
field energy and the total energy of the system respectively. Such
higher order estimates are needed to keep control of products of
creation or annihilation operators. In particular they are
essential in the proof of Theorem
\ref{theorem:many_photon_states_nonrel}. We begin by showing the
boundedness of $H_f(H+i)^{-1}$. This will give us the first step
in the inductive proof of Lemma
\ref{lemma:estim_higher_ord_nonrel}.

\begin{lemma}\label{lemma:H_f_H_bound}
The operator $H_f (H+i)^{-1}$ is bounded.
\end{lemma}
\begin{proof}
We write
\begin{equation}\label{eq:[H_f,H^1/2]}
\begin{split}
H_{f} (H+i)^{-1} = \; & (H+i)^{-1/2} H_{f} (H+i)^{-1/2} \\ &+ \,
[H_{f} , (H+i)^{-1/2}] \, (H+i)^{-1/2} .
\end{split}
\end{equation}
The first term on the r.h.s. of (\ref{eq:[H_f,H^1/2]}) is bounded,
because $H_{f}^{1/2} (H+i)^{-1/2}$ is bounded. To show that also
the second term on the r.h.s. of (\ref{eq:[H_f,H^1/2]}) is
bounded, we use the expansion
\begin{equation*}
(H+i)^{-1/2} = \frac{1}{\pi} \int_0^{\infty} dy \,
\frac{1}{\sqrt{y}} \, \frac{1}{y+H+i},
\end{equation*}
which yields
\begin{equation*}
\begin{split}
[H_{f} , &(H+i)^{-1/2}] \, (H+i)^{-1/2} \\ = \; &\frac{1}{\pi}
\int_{0}^{\infty} dy \, \frac{1}{\sqrt{y}} \, \frac{1}{y+H+i} [H ,
H_f ] \frac{1}{y+H+i} (H+i)^{-1/2} \\ = \; &\frac{i}{2 \pi}
\sum_{j=1}^{N} \int_{0}^{\infty} dy  \, \frac{1}{\sqrt{y}} \,
\frac{1}{y+H+i}\{ (p_j + A (x_j)) \cdot \phi ( i \omega G_{x_{j}})
+ h.c.\} \\ &\times  \frac{1}{y+H+i} (H+i)^{-1/2}.
\end{split}
\end{equation*}
Here and henceforth we use the notation $\phi (h) = 1 / \sqrt{2} (
a(h) + a^{*} (h))$. The r.h.s. of the last equation is clearly a
bounded operator, because $\| (y+H+i)^{-1} (p_j + A (x_j)) \|$ and
$\| \phi (\omega G_{x_{j}})(H+i)^{-1/2} \|$ are bounded by some
constants, while the second factor $(y+H+i)^{-1}$, which has norm
less than $((y-c)^2 +1)^{-1/2}$, for some constant $c >0$, ensures
absolute convergence of the integral.
\end{proof}

Now we are ready to prove Lemma
\ref{lemma:estim_higher_ord_nonrel}.

{\em Proof of Lemma \ref{lemma:estim_higher_ord_nonrel}.} We
proceed by induction over $n$. For $n=1$ the boundedness of i) is
trivial and that of ii) follows from Lemma
\ref{lemma:H_f_H_bound}. Next we assume that the statement of the
Lemma holds for positive integers less or equal to a given $n$ and
prove it for $n+1$.

First note that the boundedness of ii) follows from that of i),
since
\begin{equation*}
\begin{split}
H_{f}^{n+1} (H+i)^{-n-1} = \; & H_{f} (H+i)^{-1} H_{f}^{n}
(H+i)^{-n}  \\ &- H_f (H+i)^{-1} [ H_{f}^{n} , H] (H+i)^{-n-1}.
\end{split}
\end{equation*}
So it is enough to show the boundedness of the operator i) for $n$
replaced by $n+1$. To this end we write, using the commutator
expansion (\ref{eq:comm_exp1}),
\begin{equation}\label{eq:comm_esp}
\begin{split}
[H_{f}^{n} , H ] &(H+i)^{-n-1} =  \frac{1}{2} \, \sum_{j=1}^{N} \,
[ H_{f}^{n} , (p_{j} + A (x_{j}) )^2 ] \, (H+i)^{-n-1} \\
&=\frac{1}{2} \, \sum_{j=1}^{N} \, \sum_{l=1}^{n} \, \left(
\begin{array}{c} n \\ l
\end{array}\right) \ad_{H_f}^{l} ((p_j + A (x_j ) )^2) H_{f}^{n-l} (H+i)^{-n-1}.
\end{split}
\end{equation}
Now, using $[H_f , \phi (h) ] = -i \phi (i \omega h)$, we find,
using the commutator expansion (\ref{eq:comm_exp2})
\begin{equation}\label{eq:ad_l}
\begin{split}
\ad_{H_{f}}^{l} ((p_{j} + A (x_{j}))^2) = \; &(-i)^{l} \,
\sum_{m=1}^{l-1} \, \left( \begin{array}{c} l \\ m
\end{array}\right) \phi (i^m \omega^{m} G_{x_{j}}) \cdot \phi (i^{l-m}
\omega^{l-m} G_{x_{j}}) \\ &+ 2 \, (-i)^{l} \, \phi (i^l
\omega^{l} G_{x_{j}}) \cdot (p_{j}+A (x_{j})) \\ &+ (-i)^{l} \,
\sum_{k=1}^{3} \, [ (p_{j,k} + A_{j,k}) , \phi (i^l \omega^l
G_{x_{j},k})].
\end{split}
\end{equation}
If we insert the terms in the sum over $m$ on the r.h.s. of
(\ref{eq:ad_l}) into the r.h.s. of (\ref{eq:comm_esp}) we find
contributions to $ [H_{f}^{n} , H] (H+i)^{-n-1}$ which are
proportional to
\begin{equation*}
\begin{split}
\phi (i^m \omega^{m} G_{x_{j}}) \cdot &\phi (i^{l-m} \omega^{l-m}
G_{x_{j}}) H_{f}^{n-l} (H+i)^{-n-1} \\ = \; &\left\{ \phi \right.
(i^m \omega^{m} G_{x_{j}}) \cdot \phi (i^{l-m} \omega^{l-m}
G_{x_{j}}) (H_f + 1)^{-1}\left.  \right\} \\ &\times \left\{ (H_f
+ 1)
 H_{f}^{n-l} (H+i)^{-n-1} \right\} .
\end{split}
\end{equation*}
These contributions are all bounded. To see this note that the
operator in the first braces on the r.h.s. of the last equation is
bounded, by Lemma \ref{lemma:ann_cre_bound}, and that the operator
in the second braces is bounded, too, by induction assumption.
Similarly we can see that if we insert the terms in the sums over
$k$ on the r.h.s. of (\ref{eq:ad_l}) into the r.h.s. of
(\ref{eq:comm_esp}) we find contributions to $[H_{f}^{n} , H]
(H+i)^{-n-1}$ which are bounded. Finally, if we insert the second
term on the r.h.s. of (\ref{eq:ad_l}) into the r.h.s. of
(\ref{eq:comm_esp}), we find a contribution proportional to
\begin{equation}\label{eq:p_A_contr}
\begin{split}
\phi (i^l \omega^{l} & G_{x_{j}}) \cdot (p_{j}+A
(x_{j}))H_{f}^{n-l} (H+i)^{-n-1}
\\
= \; &\phi  (i^l \omega^{l} G_{x_{j}}) (H_f +d)^{-1} (p_{j}+A
(x_{j})) (H_{f}+d) H_{f}^{n-l} (H+i)^{-n-1} \\ &+ \text{bounded} ,
\end{split}
\end{equation}
where we inserted the operator $(H_f+ d)^{-1} (H_f+d)$ on the left
side, between the operators $\phi (i^l \omega^{l} G_{x_{j}})$ and
$(p_{j}+A (x_{j}))$, and then commuted the factor $(H_f+d)$ to the
right of $(p_{j}+A (x_{j}))$. The contribution due to the
commutator of these two terms is bounded. Now we commute one of
the $n+1$ resolvents $(H+i)^{-1}$ to the left of the field
Hamiltonians, in order to bound the factor $(p_j + A (x_j))$. The
term on the r.h.s. of (\ref{eq:p_A_contr}) becomes then
\begin{equation}\label{eq:p_A_contr2}
\begin{split}
\phi (i^l \omega^{l}  G_{x_{j}}) &(H_f +d)^{-1} (p_{j}+A (x_{j}))
(H_{f}+d) H_{f}^{n-l} (H+i)^{-n-1}\\ = \; \phi (i^l &\omega^{l}
G_{x_{j}}) (H_f +d)^{-1} (p_{j}+A (x_{j})) (H+i)^{-1} \\ \times \{
&(H_f +d) H_f^{n-l} (H+i)^{-n} + d \, [ H_{f}^{n-l}, H ]
(H+i)^{-n-1} \\ &+ [ H_{f}^{n-l+1}, H ] (H+i)^{-n-1} \}.
\end{split}
\end{equation}
For $l > 1$ the term on the r.h.s. of (\ref{eq:p_A_contr2}) is
bounded, by induction hypothesis. For $l=1$ the operator
corresponding to the sum of the first two terms in the braces on
the r.h.s. of (\ref{eq:p_A_contr2}) is also bounded by induction
hypothesis. It follows, from (\ref{eq:comm_esp}) and
(\ref{eq:ad_l}), that
\begin{equation*}
\begin{split}
[H_{f}^{n} , H] (H+i)^{-n-1} = \; &- i  \sum_{j=1}^{N} \phi (i
\omega G_{x_{j}}) (H_{f} +d)^{-1} (p_j +A (x_{j}))(H+c)^{-1} \\
&\times [H_{f}^{n} , H] (H+i)^{-n-1} \, + \, \text{bounded}.
\end{split}
\end{equation*}
If we define $A_d$ to be the operator to the left of $[H_{f}^{n} ,
H] (H+i)^{-n-1}$ in the first term on the r.h.s. of last equation,
we have
\begin{equation*}
(1 - A_d )[H_{f}^{n} , H] (H+i)^{-n-1} = \text{bounded}.
\end{equation*}
Since the norm of $\phi (i \omega G_{x_{j}}) (H_{f} +d)^{-1}$ can
be made arbitrarly small, by choosing $d$ sufficiently large, it
follows that $\| A_d \| < 1 $ for suitable $d$, and thus also the
operator $[H_{f}^{n} , H] (H+i)^{-n-1}$ has to be bounded.
\begin{flushright}$\square$
\end{flushright}

\section{Pseudo--relativistic QED: Higher Order Estimates.}
\label{sec:est_high_ord_rel} In this section we want to prove
Lemma \ref{lemma:est_higher_ord_halbrel}, which is the analogous
of Lemma \ref{lemma:estim_higher_ord_nonrel} for the
pseudo--relativistic model. This Lemma ensures the boundedness of
higher powers of the field Hamiltonian $H_f$ w.r.t. higher powers
of the total Hamiltonian $H$. Then we will apply this lemma to
prove Theorem \ref{theorem:many_phot_states_halbrel}, which
establishes the existence of asymptotic states with an arbitrary
number of free photons. Also here, as in the non--relativistic
case, we begin by showing the boundedness of $H_f (H+i)^{-1}$.
This result will be used as first step in the inductive proof of
Lemma \ref{lemma:est_higher_ord_halbrel}.
\begin{lemma}\label{lemma:H_f_H_bound_halbrel}
The operator $H_{f} (H+i)^{-1}$ is bounded.
\end{lemma}
\begin{proof}
We use
\begin{equation*}
\begin{split}
H_{f}^{2} &= \sum_{\lambda} \int dk \, |k| a_{\lambda}^{*} (k)
a_{\lambda} (k) H_{f} \\ &= \sum_{\lambda} \int dk \, |k|
a_{\lambda}^{*} (k) (H_{f}+|k| ) a_{\lambda} (k)
\end{split}
\end{equation*}
to write
\begin{equation}\label{eq:first}
\| H_{f} (H+ c)^{-1} \psi \|^{2} = \sum_{\lambda} \int dk \, |k|
\, \| (H_{f} + |k|)^{1/2} a_{\lambda} (k) (H+c)^{-1} \psi \|^{2} .
\end{equation}
Now we apply the pull--through--formula
\begin{equation*}
\begin{split}
a_{\lambda} (k) (H+c)^{-1} = \; & (H+|k|+c)^{-1} a_{\lambda} (k)
\\ &+ (H + |k| + c)^{-1} \, [ \Omega , a_{\lambda} (k) ] \,
(H+c)^{-1} ,
\end{split}
\end{equation*}
which can be easily proved by commuting the Hamiltonian $H$ with
$a_{\lambda} (k)$, and we find, from (\ref{eq:first})
\begin{equation}\label{eq:one_plus_two}
\begin{split}
\| H_{f} (H+c)^{-1} \psi \|^{2} \leq \; 2 \sum_{\lambda} \int dk
\, |k| \| (H_{f} +|k|)^{1/2} (H+|k| +c)^{-1} a_{\lambda} (k) &\psi
\|^{2} \\ + 2 \sum_{\lambda} \int dk \, |k|  \| (H_{f} +
|k|)^{1/2} (H + |k| +c)^{-1} \, [\Omega , a_{\lambda} (k) ] (H +
c)^{-1} &\psi \|^{2} .
\end{split}
\end{equation}
The first term on the r.h.s. of (\ref{eq:one_plus_two}) is bounded
by $2 \| \psi \|^{2}$, for $c>0$ sufficiently large. To see this,
note that
\begin{equation}\label{eq:bound_of_one}
\begin{split}
\| (H_{f} + |k|)^{1/2} &(H + |k| +c)^{-1} a_{\lambda} (k) \psi
\|^{2} \\ &= \sprod{\psi}{a_{\lambda}^{*} (k) (H + |k| +c)^{-1}
(H_{f} + |k|) (H + |k| +c)^{-1} a_{\lambda} (k) \psi} \\ &\leq
\sprod{\psi}{a_{\lambda}^{*} (k) (H_{f} + |k| +1)^{-1} a_{\lambda}
(k) \psi} \\ &\leq \sprod{\psi}{a_{\lambda}^{*} (k) a_{\lambda}
(k) (H_{f} +1)^{-1} \psi} ,
\end{split}
\end{equation}
where we used that $( H_{f} + |k| ) \leq (H + |k| +c)$ and that
$(H +|k| + c)^{-1} \leq (H_{f} + |k|+1)^{-1}$ for $c>0$
sufficiently large. From (\ref{eq:bound_of_one}), after
integration over $k$ and sum over $\lambda$, it follows that the
first term on the r.h.s. of (\ref{eq:one_plus_two}) is bounded by
$2 \, \| \psi \|^{2}$.

We consider now the second term on the r.h.s. of
(\ref{eq:one_plus_two}). We have
\begin{equation}\label{eq:bound_of_two}
\begin{split}
\| (H_{f} + |k|)^{1/2} &(H + |k| +c)^{-1} \, [\Omega , a_{\lambda}
(k) ] (H + c)^{-1} \psi \| \\ \leq \; & \| (H_{f} + |k|)^{1/2} (H
+ |k| +c)^{-1/2} \| \, \|(H + |k| +c)^{-1/2} \Omega^{1/2}\| \\
&\times \| \Omega^{-1/2} [\Omega , a_{\lambda} (k) ] \Omega^{-1/2}
\| \, \| \Omega^{1/2} (H + c)^{-1}\| \, \| \psi \| \\ \leq \;
&\const \, \| \Omega^{-1/2} [\Omega , a_{\lambda} (k) ]
\Omega^{-1/2} \| \, \| \psi \|
\end{split}
\end{equation}
for all $k$ and $\lambda$. Now we expand the commutator in the
term on the r.h.s. of the last equation, using Lemma
\ref{lemma:commutator}, and we get
\begin{equation}\label{eq:bound_of_comm}
\begin{split}
\| \Omega^{-1/2} [\Omega , a_{\lambda} (k) ] \Omega^{-1/2} \|
\leq \; &\frac{\sqrt{2}}{\pi} \int_{1}^{\infty} dy \sqrt{y-1} \|
(y+ (p+A)^{2})^{-1} \|^{2}  \\ &\times \| \Omega^{-1/2} (p+A)
\cdot G_{\lambda , x} (k) \Omega^{-1/2} \| \\ \leq \; &\const \,
\| \Omega^{-1/2} (p+A) \cdot G_{\lambda , x} (k) \Omega^{-1/2} \|
\\ \leq \; &\const \, \frac{\kappa (k)}{\sqrt{|k|}} \, \|
\Omega^{-1/2} (p+A) e^{-i k\cdot x} \Omega^{-1/2} \| .
\end{split}
\end{equation}
After commutation of the exponential $e^{-i k\cdot x}$ in the term
on the r.h.s. of (\ref{eq:bound_of_comm}) to the right of the
factor $\Omega^{-1/2}$ we find
\begin{equation*}
\| \Omega^{-1/2} (p+A) e^{-i k\cdot x} \Omega^{-1/2} \| \leq
\const (1 + |k| + |k|^{2}) .
\end{equation*}
Inserting this into the r.h.s. of (\ref{eq:bound_of_comm}), the
resulting bound into the r.h.s. of (\ref{eq:bound_of_two}) and
then in (\ref{eq:one_plus_two}) we find, after integration over
$k$ (which gives no problem because of the UV--cutoff), and after
sum over $\lambda$, that also the second term on the r.h.s. of
(\ref{eq:one_plus_two}) is bounded by $C \, \| \psi \|^{2}$, for
some finite constant $C > 0$. This proves the lemma.
\end{proof}

We are now ready to prove Lemma
\ref{lemma:est_higher_ord_halbrel}. The ideas of the proof are the
same as in the non-relativistic case but the estimates are more
involved because of the complicated form of the interaction
between photons and electron.

{\em Proof of Lemma \ref{lemma:est_higher_ord_halbrel}.} We
proceed by induction over $n$. For $n=1$ the boundedness of i) is
trivial and the boundedness of ii) follows from Lemma
\ref{lemma:H_f_H_bound_halbrel}. Now assume that the statement of
the lemma holds true for any integer $m$ less or equal to a given
$n$. We prove the statement for $m=n+1$. As in the
non-relativistic case it suffices to prove the boundedness of the
operator i). To this end we use the expansion given in Lemma
\ref{lemma:commutator} and the commutator expansion
(\ref{eq:comm_exp1}) to write
\begin{equation}\label{eq:comm_esp_rel}
\begin{split}
[&H_{f}^{n} , H ] (H+i)^{-n-1} =  \, [ H_{f}^{n} , \Omega ] \,
(H+i)^{-n-1} \\ &=\frac{-1}{\pi} \, \int_{1}^{\infty} dy \,
\sqrt{y-1} \,  [H_{f}^{n} , (y + (p+A)^2 )^{-1} ] \, (H+i)^{-n-1}
\\ &= \frac{-1}{\pi} \, \sum_{l=1}^{n} \left( \begin{array}{c} n
\\ l
\end{array}\right) \int_{1}^{\infty} dy \, \sqrt{y-1} \, \ad_{H_f}^{l} ((y + (p+A)^{2})^{-1})
H_{f}^{n-l} (H+i)^{-n-1}.
\end{split}
\end{equation}
We set $ \Gamma = (y+ (p+A)^2)^{-1}$ and $\Lambda_{m} =
\ad_{H_f}^{m} ((p+A)^2)$. Now we note that each factor
$\ad_{H_f}^{l} ((y + (p+A)^{2})^{-1})$ on the r.h.s. of
(\ref{eq:comm_esp_rel}) can be expanded in a sum of terms like
\begin{equation*}
\const \, \Gamma \Lambda_{m_1} \Gamma \dots \Gamma \Lambda_{m_r}
\Gamma , \quad \text{with} \quad m_i \in \{ 1, \dots , l \}, \quad
m_1 + \dots +m_r = l .
\end{equation*}
Using
\begin{equation*}
\begin{split}
\ad_{H_{f}}^{l} ((p + A (x))^2) = \; &(-i)^{l} \, \sum_{m=1}^{l-1}
\, \left( \begin{array}{c} l \\ m
\end{array}\right) \phi (i^m \omega^{m} G_{x}) \cdot \phi (i^{l-m}
\omega^{l-m} G_{x}) \\ &+ 2 \, (-i)^{l} \, \phi (i^l \omega^{l}
G_{x}) \cdot (p+A (x)) \\ &+ (-i)^{l} \, \sum_{k=1}^{3} \, [
(p_{k} + A_{k}) , \phi (i^l \omega^l G_{x,k})],
\end{split}
\end{equation*}
we see that the r.h.s. of (\ref{eq:comm_esp_rel}) can be written
as a sum of terms like
\begin{equation}\label{eq:gamma_prime}
\const \, \int_{1}^{\infty} dy \, \sqrt{y-1} \, \Gamma
\Lambda_{1}^{\prime} \Gamma \dots \Gamma \Lambda_{r}^{\prime}
\Gamma H_{f}^{n-l} (H+i)^{-n-1},
\end{equation}
with $1 \leq r \leq l$, $l \in \{1, \dots , n \}$, and where each
$\Lambda_{i}^{\prime}$ is either the product of two
field--operators, as $a^{\sharp} (i^j \omega^j G_{x}) \cdot
a^{\sharp} (i^{s} \omega^{s} G_{x})$, or the product of a
field--operator with a factor $(p+A)$, as $a^{\sharp} (i^m
\omega^m G_{x}) \cdot (p+A)$ or $(p+A)\cdot a^{\sharp} (i^m
\omega^m G_{x})$. We consider first the terms like
(\ref{eq:gamma_prime}) with $r \geq 2$. To this end expand each
field operator $a^{\sharp} ( i^j \omega^j G_{x})$ in the
$\Lambda_i^{\prime}$, writing
\begin{equation*}
a^{*} ( i^j \omega^j G_{x}) = i^j \sum_{\lambda} \int dk |k|^{j}
\frac{ \kappa (k)}{\sqrt{|k|}} \epsilon_{\lambda} (k) e^{-i k
\cdot x} a_{\lambda}^{*} (k),
\end{equation*}
and similarly for the annihilation operators. Next commute each
operator--valued distribution $a_{\lambda}^{\sharp} (k)$ to the
right of all the resolvents $\Gamma$, using the commutation
relations
\begin{equation*}
\begin{split}
[a_{\lambda}^{\sharp} (k) , (p+A) ] &= \pm \frac{1}{\sqrt{2}} \,
G_{\lambda , x}^{\sharp} (k)  \\ [a_{\lambda}^{\sharp} (k) ,
\Gamma ] &= \pm \frac{1}{\sqrt{2}} \, \Gamma \left\{ (p+A) \cdot
G_{\lambda , x}^{\sharp} (k) + G_{\lambda , x}^{\sharp} (k) \cdot
(p+A) \right\} \Gamma .
\end{split}
\end{equation*}
At the end we can write each operator like (\ref{eq:gamma_prime}),
with $r \geq 2$, as a sum of terms like
\begin{equation}\label{eq:delta}
\begin{split}
\const \, \int_{1}^{\infty} dy \sqrt{y-1} \, &\sum_{\lambda_1 ,
\dots , \lambda_m} \int dk_1 \dots dk_m \, \frac{\kappa (k_1)
\dots \kappa (k_m)}{\sqrt{ |k_1 | \dots |k_m| }} \\ &\times \Gamma
\delta_1 \Gamma \dots \Gamma \delta_{r^{\prime}} \Gamma
a_{\lambda_1}^{\sharp} (k_1) \dots a_{\lambda_m}^{\sharp} (k_m)
H_f^{n-l} (H+i)^{-n-1},
\end{split}
\end{equation}
where $r^{\prime} \geq 2$ (since we started with $r\geq 2$ and by
the commutations of the $a_{\lambda}^{*} (k)$ the number of
resolvents $\Gamma$ could only get larger), $m \leq 2l$ (since in
(\ref{eq:gamma_prime}) we had at most $2r \leq 2l$ operators
$a^{\sharp} (i \omega G)$ and since by the commutations of the
$a_{\lambda}^{*}(k)$ the number of such fields could only get
smaller), and where each operator $\delta_i$ is either a bounded
operator or the product of a factor $(p+A)$ with a bounded
operator. In both cases we have $ \| \Gamma^{1/2} \delta_i
\Gamma^{1/2} \| \leq \const \, 1 / \sqrt{y}$, and thus
\begin{equation*}
\| \Gamma \delta_1 \Gamma \dots  \Gamma \delta_{r^{\prime}} \Gamma
\| \leq \const \, \frac{1}{y^{\frac{r^{\prime}}{2}+1}}.
\end{equation*}
It follows that each term like (\ref{eq:delta}) has norm bounded
by
\begin{equation}\label{eq:bound_of_delta}
\begin{split}
\const &\int_{1}^{\infty} dy \,
\frac{\sqrt{y-1}}{y^{\frac{r^{\prime}}{2}+1}} \: \{ \sum_{
\lambda_1 , \dots , \lambda_m} \int dk_1 \dots dk_m \frac{\kappa
(k_1) \dots \kappa (k_m)}{\sqrt{ |k_1 | \dots |k_m| }}  \\
&\times \| a_{\lambda_1}^{\sharp} (k_1) \dots
a_{\lambda_m}^{\sharp} (k_m) (H_{f} + 1)^{-l} \|  \}  \: \| (H_f
+1)^{l} H_f^{n-l} (H+i)^{-n-1} \|.
\end{split}
\end{equation}
By Lemma \ref{lemma:ann_cre_bound}, and because we have at most
$2l$ operators $a_{\lambda}^{\sharp}(k)$, we find, that the
integrals and the sums inside the braces are bounded by some
finite constant. Moreover, since $r^{\prime} \geq 2$, the
integration over $y$ yields another finite constant. Finally the
factor $\| (H_f +1)^{l} H_f^{n-l} (H+i)^{-n-1} \|$ is bounded by
induction hypothesis. It follows that (\ref{eq:bound_of_delta}) is
bounded by a finite constant. We want now to consider the terms
like (\ref{eq:gamma_prime}) with $r=1$. These terms are of the
form
\begin{equation}\label{eq:r=1}
\const \int_{1}^{\infty} dy \sqrt{y-1} \Gamma \Lambda \Gamma
H_{f}^{n-l} (H+i)^{-n-1},
\end{equation}
where again $\Lambda$ is either the product of two
fields--operators or the product of a field operator with a factor
$(p+A)$. If $\Lambda$ is the product of two fields $a^{\sharp}(i^j
\omega^j G_{x})$ then we can proceed as before, expanding the
fields in integrals of operator--valued distributions
$a_{\lambda}^{\sharp} (k)$, commuting these distributions to the
right, and using the two resolvents $\Gamma$ to ensure the
convergence of the integral over $y$. It remains to consider the
case where $\Lambda$ is the product of a field $a^{\sharp}(i^j
\omega^j G_{x})$ with a factor $(p+A)$. We consider the case
$\Lambda = a^{*}(i^j \omega^j G_{x}) \cdot (p+A)$, the other
cases, with $a^{*}$ and $(p+A)$ interchanged, and with an
annihilation operator $a$ instead of $a^{*}$, being similar. To
this end we expand the field $a^{*}(i^j \omega^j G_{x})$ in an
integral over $k$ and we commute the factor $a_{\lambda}^{*} (k)
(p +A )$ to the right of the resolvent. This yields
\begin{equation}\label{eq:comm_of_field}
\begin{split}
\const &\int_{1}^{\infty} dy \, \sqrt{y-1} \, \Gamma a^{*}(i^j
\omega^j G_{x}) \cdot (p+A)\Gamma H_{f}^{n-l} (H+i)^{-n-1}\\ = \;
&\const \int_{1}^{\infty} dy \sqrt{y-1} \, \sum_{\lambda} \int dk
\, \frac{\kappa (k)}{\sqrt{|k|}} \epsilon_{\lambda , j} (k) \,
\Gamma \, e^{-ik\cdot x} \,
 \Gamma \\
 &\times a_{\lambda}^{*}(k)  (p+A )H_{f}^{n-l}
(H+i)^{-n-1}  \\ &+ \text{bounded ,}
\end{split}
\end{equation}
because the commutator of $\Gamma$ with $a_{\lambda}^{*} (k) (p +
A)$ gives factors which can be handled as we did with the terms
with $r >1$ (and whose norm is hence bounded). Now we insert, in
the operator on the r.h.s. of (\ref{eq:comm_of_field}), between
the factors $a_{\lambda}^{*}(k)$ and $(p +A)$, the operator $(H_f
+d)^{-1} (H_f +d)$, for some $d > 0$, and then we commute the term
$(H_f +d)$ to the right of $(p +A)$. The operator on the r.h.s. of
(\ref{eq:comm_of_field}) can then be written as
\begin{equation}\label{eq:comm_of_H_f}
\begin{split}
\const &\int_{1}^{\infty} dy \sqrt{y-1} \, \sum_{\lambda} \int dk
\, \frac{\kappa (k)}{\sqrt{|k|}} \epsilon_{\lambda , j} (k) \,
\Gamma \, e^{-ik\cdot x} \,
 \Gamma \\ &\times a_{\lambda}^{*}(k) (H_{f}+d)^{-1} (p_{j}+A_{j} )  (H_{f}+d)
 H_{f}^{n-l} (H+i)^{-n-1}  \\
&+ \text{bounded,}
\end{split}
\end{equation}
because the commutator of $(H_f + d)$ with the factor $(p + A)$
gives only another field $\phi (i \omega G_{x})$ which can be
bound by another resolvent $(H_f + 1)^{-1}$. Finally we commute
one of the $n+1$ resolvents $(H+i)^{-1}$ to the left of the field
Hamiltonians in order to bound the term $(p + A )$. The operator
in (\ref{eq:comm_of_H_f}) is then equal to
\begin{equation}\label{eq:last_step}
\begin{split}
\const  &\int_{1}^{\infty} dy \sqrt{y-1} \, \sum_{\lambda} \int dk
\, \frac{\kappa (k)}{\sqrt{|k|}} \epsilon_{\lambda , j} (k) \,
\Gamma \, e^{-ik\cdot x} \,
 \Gamma \, a_{\lambda}^{*}(k) (H_{f}+d)^{-1} \\
 &(p_{j}+A_{j} ) (H+i)^{-1} \{ (H_f +d) H_f^{n-l} (H+i)^{-n}  \\
 &+ d \, [ H_{f}^{n-l} , H]
 (H+i)^{-n-1} + [H_f^{n-l+1} , H] (H+i)^{-n-1} \}.
\end{split}
\end{equation}
If $l>1$ the operator in (\ref{eq:last_step}) is bounded by
induction hypothesis, since the presence of two factors $\Gamma$
ensures the absolute convergence of the integral over $y$. If
$l=1$ the contributions arising from the first two terms inside
the braces are bounded, too, by induction hypothesis. The only
contributions which could be unbounded are those arising from the
third term in the braces in (\ref{eq:last_step}) if $l=1$. Adding
all these potentially unbounded contributions together (such
contributions arise from terms like (\ref{eq:r=1}), with $l=1$ in
both cases, wheater $\Lambda$ contains a creation or an
annihilation operator), we find, from (\ref{eq:comm_esp_rel}),
\begin{equation*}
\begin{split}
[H_{f}^n , H] &(H+i)^{-n-1} = \const \int_{1}^{\infty} dy
\sqrt{y-1} \, \sum_{\lambda} \int dk \, \frac{\kappa
(k)}{\sqrt{|k|}} \epsilon_{\lambda , j} (k) \, \Gamma \,
e^{-ik\cdot x} \,
 \Gamma \\ &\times ( a_{\lambda}^{*}(k) + a_{\lambda} (k) ) (H_{f}+d)^{-1}
 (p_{j}+A_{j} ) (H+i)^{-1}  [
 H_{f}^n , H] (H+i)^{-n-1} \\
 &+ \text{bounded}.
\end{split}
\end{equation*}
Now we define $A_d$ to be the operator to the left of $[H_{f}^n ,
H] (H+i)^{-n-1}$ in the first term on the r.h.s. of last equation.
Then it follows
\begin{equation*}
(1 -A_d ) \, [H_{f}^n , H] (H+i)^{-n-1} = \text{bounded.}
\end{equation*}
Since the norm of $\phi (i \omega G_{x_{j}}) (H_{f} +d)^{-1}$ can
be made arbitrarly small, by choosing $d$ sufficiently large, it
follows that $\| A_d \| < 1 $ for suitable $d$, and thus, by the
last equation,  also the operator $[H_{f}^{n} , H] (H+i)^{-n-1}$
has to be bounded.
\begin{flushright}$\square$
\end{flushright}

With the help of Lemma \ref{lemma:est_higher_ord_halbrel}, we can
now give the proof of Theorem
\ref{theorem:many_phot_states_halbrel}.

{\em Proof of Theorem \ref{theorem:many_phot_states_halbrel}.} We
first assume that $h_i \in \mathcal{C}_{0}^{\infty} (\R^3
\backslash \{ 0 \} )$, for all $i \in \{ 1, \dots n \}$, and that
$\ph = \chi (H \leq E ) \ph $ for some $E < \infty$. We prove then
the limit (\ref{eq:limit_many_rel}) by induction over $n$. For
$n=1$, (\ref{eq:limit_many_rel}) follows by Theorem
\ref{theorem:defandprop}. Assume now that
(\ref{eq:limit_many_rel}) holds true for any integer less than a
given $n$. We want to  prove it for $n$ fields $a_{+}^{\sharp}
(h_i)$. We consider the case where we have $n$ creation operators:
the other cases are then similar. Since $\ph = \chi (H \leq E )
\ph $, since the functions $h_i$ have compact support, and because
of Theorem \ref{theorem:defandprop}, iv), the vector $a_{+}^{*}
(h_{l}) \dots a_{+}^{*} (h_{n}) \ph$ is well defined, and
\begin{equation}\label{eq:a+l_n_rel}
a_{+}^{*} (h_{l}) \dots a_{+}^{*} (h_{n}) \ph = \lim_{t \to
\infty} e^{iHt} a^* (h_{l,t})e^{-iHt} a_{+}^{*} (h_{l+1}) \dots
a_{+}^{*} (h_{n}) \ph ,
\end{equation}
for each $l\in \{1,\dots,n\}$. To show that the limit
(\ref{eq:limit_many_rel}) holds true we consider now the
difference
\begin{equation}\label{eq:trenn_rel}
\begin{split}
e^{iHt} a^{*} (h_{1,t}) \dots &a^{*} (h_{n,t}) e^{-iHt}\ph -
a_{+}^{*} (h_{1} ) \dots a_{+}^{*} (h_{n}) \ph \\ = \; &(e^{iHt}
a^{*} (h_{1,t})e^{-iHt} - a_{+}^{*} (h_1)) a_{+}^{*} (h_2) \dots
a_{+}^{*}(h_n) \ph \\ &+ e^{iHt} a^{*} (h_{1,t})e^{-iHt} \times \\
&\left\{ e^{iHt} a^{*} (h_{2,t}) \dots a^{*} (h_{n,t}) e^{-iHt} -
a_{+}^{*} (h_2) \dots a_{+}^{*} (h_n) \right\} \ph .
\end{split}
\end{equation}
The norm of the first term on the r.h.s of the last equation
converges, by (\ref{eq:a+l_n_rel}), to $0$ as $t \to \infty$. To
handle the second term on the r.h.s. of (\ref{eq:trenn_rel}) we
insert the operator $ id = (H+i)^{-1} (H+i) $ just in front of the
braces, and we commute the factor $(H+i)$ through the whole
braces. The second term on the r.h.s. of (\ref{eq:trenn_rel})
becomes then
\begin{equation}\label{eq:sec_term_rel}
\begin{split}
e^{iHt} a^{*} (h_{1,t}) e^{-iHt} (H+i)^{-1} &\times \\ \left\{
(e^{iHt}  a^{*} (h_{2,t}) \dots a^{*} (h_{n,t}) \right. &e^{-iHt}
- a_{+}^{*} (h_2) \dots a_{+}^{*} (h_n))(H+i)\ph  \\ +
\sum_{l=2}^{n} (e^{iHt} a^{*} (h_{2,t}) \dots a^{*} &(\omega
h_{l,t} ) \dots a^{*} (h_{n,t}) e^{-iHt} \\ &- a_{+}^{*} (h_2)
\dots a_{+}^{*} (\omega h_l) \dots a_{+}^{*} (h_n)) \ph \\
 + e^{iHt} [\Omega ,a^{*} (h_{2,t}) \dots a^{*} &(h_{n,t})] \left. e^{-iHt}
\ph \right\} .
\end{split}
\end{equation}
Now the term in front of the braces is bounded, uniformly in $t$.
The first term inside the braces, and each factor in the sum over
$l$ converges to $0$, as $t \to \infty$, by induction hypothesis.
Hence (\ref{eq:limit_many_rel}) follows if we show that the norm
of the last term inside the braces converges to zero as $t \to
\infty$. To this end we expand the commutator with $\Omega$ in an
integral, as in Lemma \ref{lemma:commutator}. Using the
commutation relation $[a^{*} (h_t), (p+A)^{2}] = - \sqrt{2} (G_{x}
, h_t ) \cdot (p+A)$ we find
\begin{equation}
\begin{split}
\| &[\Omega ,a^{*} (h_{2,t}) \dots a^{*} (h_{n,t})] e^{-iHt} \ph
\| \\ &\leq
 \frac{\sqrt{2}}{\pi} \sum_{j=2}^{n} \int_{1}^{\infty} dy \, \sqrt{y-1} \,  \| (y +
(p+A)^{2})^{-1} \| \, \| (G_x , h_{j,t}) \| \\ &\times \| a^{*}
(h_{2,t}) \dots a^{*} (h_{j-1,t}) (p + A) a^{*} (h_{j+1,t}) \dots
a^{*} (h_{n,t}) ( y+ (p+A)^2 )^{-1} \ph_t \| ,
\end{split}
\end{equation}
where $\ph_t = e^{-iHt} \ph $. Since $ \| (G_x , h_{j,t}) \|$
converges to zero as $t \to \infty$ it only remains to show that
the vector
\begin{equation}\label{eq:vect}
a^{*} (h_{2,t}) \dots a^{*} (h_{j-1,t}) (p + A) a^{*} (h_{j+1,t})
\dots a^{*} (h_{n,t}) ( y+ (p+A)^2)^{-1} \ph_t
\end{equation}
has norm bounded by $C / y$, for some $C < \infty$, and for all $y
\geq 1$ (note that the factor $1/y$ is necessary to make the
$y$--integral absolut convergent). To this end we consider the
operator which act on $\ph_t$ in (\ref{eq:vect}) and we commute,
first of all, the factor $(p+A)$ to the very left, using the
commutation rule $[ a^{*} (h_{i,t}) , (p+A)] = - 1/\sqrt{2} \,
(G_{x} , h_{i,t})$. We find that the operator which act on $\ph_t$
in (\ref{eq:vect}) can be written as
\begin{equation}\label{eq:oper_1}
\begin{split}
( p + A ) a^{*} &(h_{2,t}) \dots  a^{*} (h_{n,t}) ( y+ (p+A)^2
)^{-1} \\ &- \frac{1}{\sqrt{2}} \, \sum_{l=2}^{j-1} (G_{x} ,
h_{l,t}) a^{*} (h_{2,t})\dots a^{*} (h_{n,t}) (y + (p+A)^2 )^{-1}.
\end{split}
\end{equation}
Next we commute, in the first term as well as in each term in the
sum in (\ref{eq:oper_1}), all the fields $a^{*} (h_{m,t})$ to the
right of the resolvent $(y + (p+A)^2 )^{-1}$. Here we use the
commutation relation
\begin{equation*}
\begin{split}
[a^{*} (h_{m,t}) , (p+A) ] = \; &- \frac{1}{\sqrt{2}} \, (G_{x} ,
h_{m,t}) \\ [a^{*} (h_{m,t}) , (y + (p+A)^{2})^{-1} ] = \; &
\sqrt{2} \, (y+(p+A)^2)^{-1} \\ &\times (p+A) \cdot (G_{x},
h_{m,t})  \, (y+(p+A)^2 )^{-1} .
\end{split}
\end{equation*}
At the end each term in the sum over $l$ in (\ref{eq:oper_1}) will
be written as a sum of terms like
\begin{equation}\label{eq:oper_2}
(y + (p+A)^2 )^{-1} \times B \times a^{*} (h_{i_1 ,t}) \dots
a^{*} (h_{ i_m ,t}),
\end{equation}
where $B$ is some bounded operator, and the number of field is at
most $n-3$. The first factor in (\ref{eq:oper_1}), on the other
hand will be written as a sum of factors like
\begin{equation}\label{eq:oper_3}
(p + A) (y + (p+A)^2 )^{-1 /2 } \times B \times (y + (p+A)^2
)^{-1} a^{*} (h_{i_1 ,t}) \dots  a^{*} (h_{ i_m ,t}),
\end{equation}
where, again $B$ is a bounded operator and $m \leq n-3$, plus the
factor
\begin{equation}\label{eq:oper_4}
(p+A) (y + (p+A)^{2})^{-1} a^{*} (h_{2 ,t}) \dots  a^{*} (h_{ n
,t}).
\end{equation}
Each term like (\ref{eq:oper_2}) or like (\ref{eq:oper_3}) gives a
contribution to the vector (\ref{eq:vect}) whose norm is bounded,
uniformly in $t$, by $C / y$, for a finite constant $C$. This
follows from the bound $\| a^{\sharp} (h_{1,t}) \dots a^{\sharp}
(h_{n ,t}) \ph_t \| \leq C$, for each $t \in \R $. This estimate
follows from Lemma \ref{lemma:ann_cre_bound}, in Appendix
\ref{sec:a_bound}, and from the boundedness of
$H_f^{n}(H+1)^{-n}$, which follows by Lemma
\ref{lemma:est_higher_ord_halbrel}.

It only remains to consider the contribution from the operator
(\ref{eq:oper_4}). To this end we commute the factor $(p+A)$
through the resolvent $(y+(p+A)^2)^{-1}$. The contribution from
the term which contains the commutator between $(p+A)$ and $(y +
(p+A)^2))^{-1}$ can be handled as we did with operators like
(\ref{eq:oper_2}), and has therefore a norm bounded by $C / y$. It
only remains to consider the contribution to the norm of the
vector (\ref{eq:vect}) arising from the term (\ref{eq:oper_4})
with $(p+A)$ and $(y+(p+A)^2 )^{-1}$ interchanged. This is bounded
by
\begin{equation}
\begin{split}
\| (y + (p+A)^{2})^{-1} &(p+A) a^{*} (h_{2 ,t}) \dots  a^{*} (h_{
n ,t}) \ph_t \|
\\ &\leq 1/y \, \| ( p+A ) a^{*} (h_{2 ,t}) \dots  a^{*} (h_{ n ,t}) \ph_t
\| \\ &\leq C / y,
\end{split}
\end{equation}
where, in the last step we used that $\| ( p+A) a^{\sharp}
(h_{1,t}) \dots a^{\sharp} (h_{n ,t}) \ph_t \| \leq C $, $\forall
t >0$, which follows, after some manipulations, from the
boundedness of the operator $[H_{f}^{n} , H ] (H +i )^{-n-1}$ (see
Lemma \ref{lemma:est_higher_ord_halbrel}). This completes the
proof of the limit (\ref{eq:limit_many_rel}), for $h_i \in
\mathcal{C}_{0}^{\infty} (\R^3 \backslash \{ 0 \})$, and for $\ph
\in \Ran \chi (H \leq E )$. The bound
(\ref{eq:bound_many_phot_halbrel}) follows now from the
corresponding bound for products of the usual fields $a^{\sharp}
(h)$. This bound permits also to show that the limit
(\ref{eq:limit_many_rel}) also holds for wave--functions $h_i \in
L_{\omega}^{2}(\R^3 ; \C^2 )$, because $\mathcal{C}_{0}^{\infty}
(\R^3 \backslash \{ 0 \})$ is dense in $L_{\omega}^{2}$. Moreover
the limit (\ref{eq:limit_many_rel}) also holds for $\ph \in
D((H+i)^{n/2})$, because $\cup_{E>0} \chi (H\leq E) \mathcal{H}$
is a core for $(H+i)^{n/2}$.
\begin{flushright}$\square$
\end{flushright}

\noindent {\bf Acknowledgements.} M.G. thanks V. Bach, M. Loss,
and M. Merkli for helpful discussions. His work was supported in
part by an award of the UAB Faculty Development Program.

\bibliographystyle{alpha}

\addcontentsline{toc}{section}{\numberline{}References}

\end{document}